# Photonic Integrated Devices for Nonlinear Optics

**Lucia Caspani,[1,\*] David Duchesne,[1,2] Ksenia Dolgaleva,[3] Sean Wagner,[3] Marcello Ferrera,[1,4] Luca Razzari,[1,5] Alessia Pasquazi,[1] Marco Peccianti,[1,6] David J. Moss,[7, 8] J. Stewart Aitchison,[3] and Roberto Morandotti[1,\*]**

[1]*INRS-EMT, 1650 Boulevard Lionel Boulet, Varennes, Québec J3X 1S2, Canada*

[2]*Massachusetts Institute of Technology, 77 Massachusetts Avenue, Cambridge, MA 02141, USA*

[3]*Edward S. Rogers Sr. Department of Electrical and Computer Engineering, University of Toronto, 10 King's College Rd, Toronto, ON M5S 3G4, Canada*

[4]*University of St Andrews, Saint Andrews, Fife KY16 9AJ, UK*

[5]*Fondazione Istituto Italiano di Tecnologia, Via Morego 30, 16163 Genova, Italy*

[6]*ISC-CNR, UOS Roma, Università 'La Sapienza', Piazzale A. Moro 2, 00185 Rome, Italy*

[7]*Institute of Photonics and Optical Science (IPOS) and CUDOS, School of Physics, University of Sydney, New South Wales 2006, Australia*

[7]*Current Address: School of Electrical and Computer Engineering, RMIT University, Melbourne, Victoria, Australia 3001*

[\*]*Corresponding authors: caspani@emt.inrs.ca; morandotti@emt.inrs.ca*

We review our recent progresses on frequency conversion in integrated devices, focusing primarily on experiments based on strip-loaded and quantum-well intermixed AlGaAs waveguides, and on CMOS-compatible high-index doped silica glass waveguides. The former includes both second- and third-order interactions, demonstrating wavelength conversion by tunable difference-frequency generation over a bandwidth of more than 100 nm, as well as broadband self-phase modulation and tunable four-wave mixing. The latter includes four-wave mixing using low-power continuous-wave light in microring resonators as well as hyper-parametric oscillation in a high quality factor resonator, towards the realization of an integrated multiple wavelength source with important applications for telecommunications, spectroscopy, and metrology.

## 1. Introduction

Optical frequency conversion has become an increasingly important area of research since the first observation of Second Harmonic Generation (SHG) by Franken and coworkers in 1961 [1], which marked the birth of nonlinear optics (see also [2,3]). After the first studies in integrated optics (for a comprehensive review of this topic see, e.g., [4]), it became clear that guided waves would offer fundamental advantages for nonlinear optics due to the intrinsic radiation confinement, thus leading to high optical intensities over long propagation distances [5-7]. Indeed, ten years later, Anderson and Boyd performed the first nonlinear optics experiment in waveguides that, as for bulk nonlinear optics, dealt with frequency conversion (SHG in GaAs waveguides) [8].

Optical frequency conversion still remains a topic of substantial interest in both bulk and integrated structures, as witnessed by the high number of published articles each year in this field [9]. Besides the obvious interest for novel wavelength sources, optical frequency conversion has attracted significant attention in the integrated optics community for applications related (but not limited) to all-optical networks, such as signal regeneration (see, *e.g.*, [10]) and Wavelength

Division Multiplexing (WDM, see, *e.g.*, [11,12] and references therein). Other applications include gas sensing, time resolved spectroscopy, high-resolution metrology, material science and bio-photonics (see [13] and references therein).

In this paper, we review our recent results on optical frequency conversion in integrated structures, exploiting $\chi^{(3)}$ effects in AlGaAs strip-loaded waveguides and in CMOS compatible high-index doped silica glass microring resonators, and the $\chi^{(2)}$ nonlinearity in AlGaAs intermixed waveguides. We focus on self-phase and cross-phase modulation, four-wave mixing, second-harmonic generation, and on the phase matching schemes required to enhance these processes. The paper is organized as follows. In Sec. 1.1 we review optical frequency conversion based on both second- and third-order optical nonlinearities. Then, we briefly review in Sec. 1.2 the photonic platforms for integrated frequency conversion, mainly focusing on semiconductors and high-index doped silica glass. In order to enhance the nonlinear interactions several photonic structures can be exploited, as described in Sec. 1.3. Our results for frequency conversion in AlGaAs strip-loaded waveguides, by means of both $\chi^{(2)}$ and $\chi^{(3)}$ interactions are reported in Sec. 2. Finally, in Sec. 3 we describe our results for frequency conversion in microring resonators, allowing for highly efficiency nonlinear interactions with continuous wave sources.

## *1.1. Frequency Conversion*

The nonlinear processes underpinning optical frequency conversion typically have electronic origin, stemming (in dielectrics) from the profile of the potential well that maintains the electron cloud localized. These processes are inherently instantaneous at optical frequencies and can be classified according to the order of the nonlinearity (typically $2^{nd}$ or $3^{rd}$), represented by the nonlinear optical susceptibilities $\chi^{(2)}$ and $\chi^{(3)}$, respectively. Third-order nonlinear optical interactions are of particular relevance for optical communication systems, considering that common materials for integrated optics like silicon and silica, being centrosymmetric, lack even-order nonlinearities. Moreover, standard interaction schemes in $\chi^{(3)}$ materials exploit the interaction of waves close in wavelength, all localized within the standard third-telecom window (C-band). Commonly used materials for optical waveguides also include second-order nonlinear media, such as AlGaAs, and thus $\chi^{(2)}$ effects may also be exploited in guided geometries. There is a large number of publications on nonlinear optical effects, both in bulk and guided optics, and a detailed overview on this topic is well beyond the scope of this paper. For a detailed and complete survey, we refer the reader to well-known books, *e.g.*, [14-16].

In the following subsections, we briefly describe some of the nonlinear processes exploited for optical frequency conversion in integrated devices, starting with $\chi^{(3)}$ effects, namely self- and cross-phase modulation and four-wave mixing. In Sec. 1.1.2 we focus instead on second-order nonlinear processes, such as second-harmonic and difference-frequency generation.

### 1.1.1. Third-order effects

*Phase modulation*. The Kerr effect [14] is the simplest model describing the nonlinear field-matter interaction experienced by intense optical beams when propagating in $\chi^{(3)}$ materials. It implies a linear dependence of the field phase velocity (*i.e.*, the refractive index) from its intensity. The *self-phase modulation* (SPM) that leads to a nonlinear variation in time of the optical phase $\phi(t)$ [17,18], is the most common phenomenon described by the Kerr-law, *i.e.*, $\phi(t) \propto n_2 I(t)$, where $n_2$ is the nonlinear Kerr coefficient (determined by the material properties) and $I(t)$ is the intensity profile of the pulse. Since the instantaneous optical frequency is proportional to the first derivative in time of the phase [14], the nonlinear phase shift due to SPM leads to spectral broadening of the pulse. SPM plays a key role in the generation of new frequencies due to the nonlinear pulse spectral broadening in third order materials, usually referred as *supercontinuum* generation [15].

When two pulses, namely *A* and *B*, overlap in time while propagating in a $\chi^{(3)}$ medium, they will also experience what is referred to as cross-phase modulation (XPM). In this case, pulse *A* undergoes a phase modulation induced by the intensity variation of pulse *B* (and *vice versa*), $\phi_{A/B}(t) \propto n_2 I_{B/A}(t)$ [15], in addition to SPM. In general, the magnitude of the Kerr coefficient of the XPM phenomenon depends on the specific interaction geometry. For example, for co-polarized pulses at close wavelengths, the XPM Kerr coefficient is twice as strong as the corresponding coefficient in SPM. In turn, XPM is inherently affected by the group velocity mismatch, leading to walk-off between the pulses.

A complete and quantitative description of SPM and XPM taking into account all the propagation effects such as dispersion, losses, group velocity dispersion and mismatch, etc., relies on the numerical integration of the propagation equations. A detailed analysis can be found in [14,15]. Nevertheless, the simple picture given above properly describes the main characteristics of SPM and XPM and the experimental results reported in the following (Sec. 2.1).

SPM and XPM have often been seen as detrimental effects in optical communication systems (*e.g.*, waveform degradation), and much effort has previously been devoted in fiber systems to avoid and limit these processes. Nevertheless, nowadays protocols exist for exploiting nonlinear phase modulations in optical communication, *e.g.* SPM-based optical regeneration of WDM channels, ultrafast all-optical switching based on both SPM and XPM, XPM-induced demultiplexing, etc. (see, *e.g.*, [19]).

***Four wave mixing***, (FWM) is the most general interaction model in third order nonlinear materials, and involves the interaction between *four distinguishable waves*, for example the generation of a new frequency component from three separate input sources [14,15]. Differently from SPM and XPM, in the FWM process, an actual energy exchange occurs amongst the interacting waves. From a quantum perspective, photons of frequencies $\omega_{p1}$, $\omega_{p2}$, and $\omega_s$ interact in a $\chi^{(3)}$ medium to produce a new photon with (energy) frequency $\omega_i = \omega_{p1} + \omega_{p2} - \omega_s$, as prescribed from energy considerations (see Fig. 1). In the degenerate case, only two separate light sources are used: $\omega_i = 2\omega_p - \omega_s$. The process can be intuitively understood as an electronic interaction between an atom and the two pump photons $\omega_p$ such that the atom is excited to a virtual level. The relaxation of this interaction is then stimulated by another incoming signal photon $\omega_s$ to allow the emission of two new photons at $\omega_s$ and $\omega_i$. Efficient frequency conversion through FWM in a medium of finite length requires the momentum conservation of the interacting waves, a condition typically reported as phase matching [14,15]. In integrated optics, this can be achieved, *e.g.*, by means of dispersion-engineered waveguides.

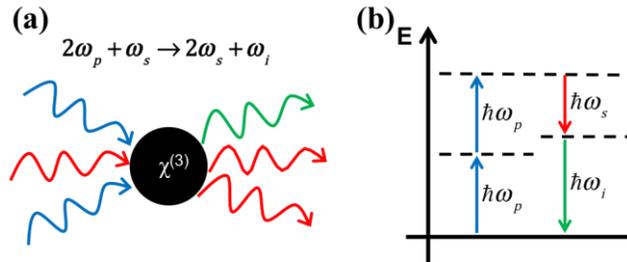

Fig. 1. (Color online) Schematic of the quantum interaction between incident and outgoing photons within the propagating nonlinear medium for a degenerate FWM process. (b) Energy diagram for degenerate FWM where the dashed lines represent virtual levels.

In telecom applications, FWM has been exploited to implement optical frequency conversion applications, such as carrier remapping in WDM systems, a result of the relaxed phase-matching condition in narrow band systems. In the following sections we report our recent results on FWM-mediated optical frequency conversion in integrated structures, both for pulsed and continuous wave (CW) sources (see Secs. 2.1 and 3.1, respectively).

**1.1.2. Second-order processes**

Similarly to FWM in $\chi^{(3)}$ materials, in second-order nonlinear media optical frequency conversion can be achieved by means of three-wave mixing (TWM). It implies an energy exchange among three distinguishable waves, governed by energy conservation (*i.e.*, $\omega_1+\omega_2=\omega_3$).

Second-harmonic generation ($\omega_1+\omega_1\rightarrow 2\omega_1$) is a particular case of sum-frequency generation (SFG, $\omega_1+\omega_2\rightarrow\omega_3$). It is widely exploited to generate short wavelengths from a single long wavelength source, *e.g.*, in frequency-doubled lasers as well as for nonlinear microscopy (two-photon fluorescence). On the other hand, difference-frequency generation (DFG, $\omega_1-\omega_2\rightarrow\omega_3$) is the main mechanism behind most optical parametric oscillators, and is also of critical importance for the generation of THz radiation, and for chemical sensing in the mid-infrared. All these effects can in principle occur, but the phase matching condition usually discriminates a dominant process among them.

In a number of relevant scenarios (*e.g.*, parametric amplifications), phase matching in second-order optical frequency conversion is typically more problematic than the equivalent case realized through FWM, since the wavelengths of the fields involved can be quite different and thus the effect of dispersion is significant. Several different techniques have been developed so far for achieving phase matching, mostly relying on birefringence in bulk nonlinear crystals. An important class of reference materials for integrated optics, namely the III-V semiconductors like AlGaAs, possesses significant second-order nonlinearity while being non-birefringent, and thus other more complex phase matching techniques must be exploited. More details on this topic will be addressed in Sec. 1.3.2.

## 1.2. Photonic Platforms

Several different material platforms have been investigated in the last years for photonic integrated circuits [20-23], and the quest is still open for the *perfect* material for future all-optical networks and applications. A non-comprehensive list includes semiconductors, such as silicon [24,25] and GaAs/AlGaAs (see, *e.g.*, [26]), as well as nonlinear glasses, such as chalcogenide [27], silicon oxynitride [28] and bismuth oxides [29].

Despite the large number of platforms investigated so far, none of them has clearly stood out as the ultimate choice for future all-optical networks, mostly due to the difficulty to address in a single material all the necessary requirements, such as low linear and nonlinear losses, high nonlinearity, mature fabrication technology, etc. Silicon, for example, features a very mature low-cost fabrication technology mainly imported from electronics, but, being an indirect-bandgap semiconductor, suffers from the difficult implementation of an electrically pumped laser. In addition, despite its remarkable nonlinear Kerr coefficient, which in principle makes it a favorable material for nonlinear applications, it is affected by significant nonlinear losses in the telecom spectral range (1400-1600nm), due to two-photon absorption first, and the subsequent absorption by the induced free-carriers [30-33].

In this paper, we focus on two of the most promising solutions currently available among semiconductors and nonlinear glasses, i.e. *AlGaAs* and *Hydex*® (a novel high-index doped silica glass developed by *Little Optics* in 2003 [34]), respectively.

### 1.2.1. AlGaAs

AlGaAs has been termed "the silicon of nonlinear optics" [26] due to its excellent nonlinear performance [26,35-38]. In addition to the highest Kerr nonlinearity among the candidates for all-optical signal processing ($n_2 \sim 10^{-17}$ m$^2$/W at $\lambda$=1.55 μm [36]), it also has a high refractive index allowing for a tight mode confinement for even more efficient nonlinear interactions. The value of the refractive index of AlGaAs can be adjusted in the range between 2.90 and 3.38 by changing the Al concentration from 100% to 0 during epitaxial growth, which offers significant freedom in designing various AlGaAs integrated optical components. Moreover, AlGaAs is a direct bandgap material that allows the monolithic integration of a laser source, low-loss waveguides (performing various linear or nonlinear operations), and a detector on the same chip without resorting to hybrid integration. All these benefits render AlGaAs a promising material platform for the realization of a broad class of integrated photonic devices.

Nonlinear effects in AlGaAs are not limited only to third-order processes. Since it is a non-centrosymmetric medium (unlike silicon), it possesses even-order nonlinearities, with a relatively large value of $\chi^{(2)}$ of about 200 pm/V [39]. Typically, the lower-order nonlinearities are stronger than the higher-order nonlinear interactions, and thus AlGaAs devices capable of SHG and DFG are regularly investigated [40].

The key to efficient second-order nonlinear wavelength conversion is to satisfy the phase matching condition. Since AlGaAs is linearly isotropic, this cannot be achieved by using birefringence as is done in bulk nonlinear crystals such as LiNbO$_3$, KTP, and BBO. Instead, other phase matching techniques must be introduced. Strong interest in exploiting the large second-order nonlinear properties of AlGaAs for the purposes of monolithic integration has spurred studies of several different structures for phase matching. A short overview of some of the main approaches is reported in Sec. 1.3.2, with a detailed description of the domain-disordered quasi-phase matching (DD-QPM) technique, that allows us to observe efficient SHG and DFG in AlGaAs waveguides (see Sec. 2.2).

**1.2.2. Hydex®**

Generally, third order nonlinear interactions encompass a multitude of parametric phenomena that all depend critically on the product of three factors [14,15,41]: (i) the *effective* interaction length *L* (or time duration) over which light interacts with the medium it propagates in, (ii) the intensity of the light beam, and (iii) the strength of the nonlinearity tensor. The latter two conditions are typically achieved using semiconductor materials [14,30,41,42] or high-index glasses such as chalcogenides [43,44], as they possess both a high nonlinear coefficient and a high index of refraction (allowing sub-wavelength confinement of the light, and thus high intensities). However, most nonlinear optical waveguide devices are typically short, ~mm, and long or resonant structures cannot be used to enhance nonlinearities due to impairments from nonlinear and linear optical losses [38,45-47]. Specifically, multiphoton absorption (a nonlinear effect related to the imaginary part of odd-order nonlinear susceptibility terms) in most semiconductors limits the overall maximum effective nonlinearity that can be achieved, and has been reported in several experiments to be detrimental for frequency mixing [38,46]. Moreover, linear losses in many optical materials remain significantly high due to the scattering sites resulting from the etching processes commonly used for the fabrication of high contrast waveguides [45,48]. Consequently, this limits the overall useful length for nonlinear interactions, and puts limitations on the maximum achievable quality factor for resonators. Illustratively shown in Fig. 2 the effective interaction length for the nonlinear gain (phase) for SPM scales as $L_{eff} = (1-\exp(-\alpha L))/\alpha$, and consequently, low losses are fundamental to exploit the length dependence of the nonlinearity. This remains a challenge for most nonlinear optical materials such as AlGaAs and silicon [38,45-47], where losses are still on the order of a few dB/cm.

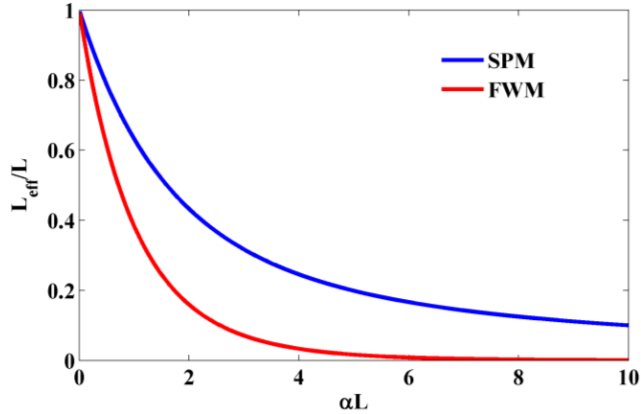

Fig. 2. (Color online) Ratio of the effective nonlinear length $L_{eff}$ to the total propagation length $L$ as a function of the linear losses. The blue curve is for the nonlinear phase accumulated for SPM, whereas the red curve is for wavelength conversion using FWM.

The Hydex® platform [34,49] was developed as a compromise between the optimal linear properties of silica glass, such as low dispersion and low linear losses, and the nonlinear properties of semiconductors: high linear and nonlinear index. It is a high-index glass (n=1.7), where buried waveguides (in silica) are formed via chemical vapor deposition and reactive ion etching [34]. The design cross-section and supported mode of Hydex® waveguides are shown in Fig. 3. Such waveguides can exhibit propagation losses as low as 0.06 dB/cm [49], a group velocity dispersion of $\beta_2 < 20$ ps$^2$/km for over 100 nm centered at 1550 nm [50], and an absence of multiphoton absorption for intensities up to 25GW/cm$^2$ (propagated over 45cm devices) [49]. Whereas the nonlinear parameter, $\gamma = 2\pi n_2/\lambda A_{eff}$, is only 220W$^{-1}$km$^{-1}$ [49] (compared with the most nonlinear silicon-on-insulator waveguides having $\gamma$ values up to 1000 times larger [51,52]), the extremely low loss associated with this material platform allows for longer waveguides (see [49] for SPM measurements in a 45 cm spiral waveguide confined on a 2×2 mm$^2$ footprint) and resonant structures having high quality factors. The importance of this optical platform has been demonstrated repeatedly in the last decade via both linear and nonlinear devices. Specifically, optical filters with more than 80 dB rejection were achieved via a cascade of high quality factor ring resonators [53]. These resonators have also been shown critically important for future generation biomolecular sensors [54], all-optical signal processing [55], and, more recently, for all-optical logical operations, such as all-optical integration [56]. For nonlinear optical applications, an ultrashort pulse compressor [57] and an ultrafast all-optical oscilloscope [58] were developed by means of a long spiral waveguide. Moreover, self-phase modulation measurements [49] and the generation of a supercontinuum [59] were recently reported, and prove the potential of utilizing this material platform for optical frequency conversion. In this paper we report experimental results showing the possibility of achieving CW FWM and spontaneous parametric oscillation (see Secs. 3.1 and 3.2, respectively) by exploiting a high quality factor microring resonator (see Fig. 3) that surpasses the limitations found in conventional nonlinear semiconductors.

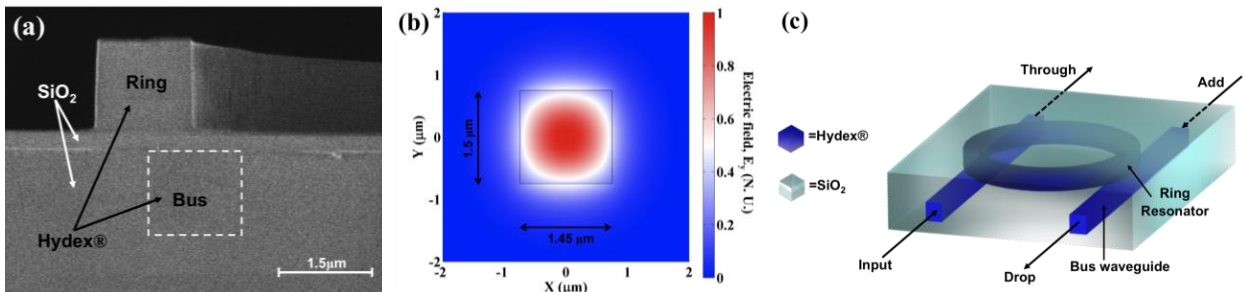

Fig. 3. (Color online) Cross-sectional view of the Hydex® waveguide showing (a) the geometry (picture of the device prior to final $SiO_2$ deposition to bury the waveguide) and (b) the distribution of the electric field of the fundamental mode. (c) Schematic of the vertically coupled microring resonator.

## 1.3. Photonic Structures

In order to achieve highly efficient optical frequency conversion, several photonic structures have been developed so far, relying on stronger mode confinement, and thus higher power density (see, *e.g.*, [30,60-62]), material dispersion compensation (see, *e.g.*, [30,63,64]), *artificial* phase matching in non-birefringent materials (see, *e.g.*, [40,64]), etc. However, these structures also present some drawbacks. For example, in photonic wires, the small size of the mode leads to scattering centers and surface state absorptions due to a relatively large field along the waveguide etched sidewalls [65]. For this reason the appropriate structure must be carefully chosen in relation with the application under investigation. In the following sections, we briefly describe the structures exploited for high-efficiency optical frequency conversion in AlGaAs and Hydex®.

### 1.3.1. AlGaAs strip-loaded waveguides

AlGaAs strip-loaded waveguides are micron-size devices that can be fabricated using standard photolithography procedures (see Fig. 4(b) for a schematic diagram). The lateral confinement of the mode in such waveguides is achieved by defining a ridge in the upper cladding, so that the guiding layer is buried underneath the ridge, and the guided mode does not experience large scattering losses due to the sidewall roughness. The waveguide dispersion in strip-loaded waveguides is very small, as the light confinement is not very tight, so, an obvious limitation to the nonlinear interactions comes from the material dispersion that introduces a phase mismatch between the interacting waves at different frequencies. Several techniques have been developed to solve the problem of phase mismatch. Among those, we can list the realization of dispersion-engineered waveguides with sub-micron dimensions (nanowires) [66] and the periodical modulation of either the linear [67] or nonlinear [68] optical susceptibility. In the next section, we discuss a few techniques for phase matching the second-order nonlinear effects.

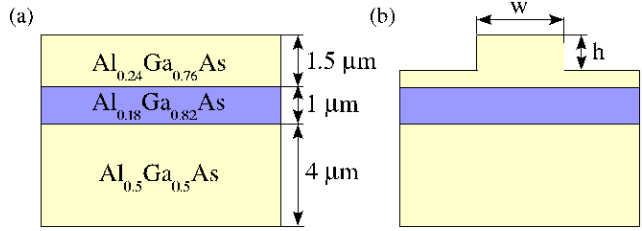

Fig. 4. (Color online) (a) The designed AlGaAs wafer composition and (b) the schematic of a strip-loaded waveguide studied in our experiments: w = 2-3μm, h=1.0-1.2μm. Image source: [47].

### 1.3.2. Periodically intermixed waveguides

The problem of phase matching second-order nonlinear processes in isotropic materials such as AlGaAs can be overcome in a number of ways. As such, several waveguide structures have been developed so far by many groups in order to achieve it in AlGaAs waveguides. These include multilayer AlGaAs/$Al_2O_3$ waveguides for artificial form birefringence phase matching (FBPM) [69,70], engineered-waveguides for modal phase matching (MPM) [71,72], Bragg reflection waveguides (BRW) for MPM [73], orientation patterned GaAs (OP-GaAs) for quasi-phase matching (QPM) via domain reversal [74], and periodically switching nonlinearity (PSN) via etch-and-regrowth for QPM [75]. A comprehensive discussion of these techniques may be found in Ref. [13,40]. Typically, these techniques require complex fabrication processes and/or waveguide structures that limit their potential for monolithic integration. For instance, the electrically insulating $Al_2O_3$ layers in FBPM methods prohibit integration of current pumped active devices such as lasers and amplifiers into the structures. Regrowth epitaxy in the QPM techniques limits device yield, increases costs, and raises optical losses. MPM methods sacrifice overlap of the interacting waves, thus reducing the conversion efficiency. In all cases, optical attenuation tends to be high due the use of high-order modes, unavoidable rough interfaces from processing, and lossy waveguiding structures. Another method for phase matching that is more amenable to monolithic integration is required in order to implement complex optical circuits for miniaturization.

One phase matching technique that has led the way for monolithic integration is domain-disordered quasi-phase matching (DD-QPM). By using simple waveguide structures and fabrication processes, DD-QPM aims to reduce optical losses while remaining cost efficient. Unlike the more traditional QPM approach, this method does not require periodically inverting the nonlinear susceptibility with complex epitaxy steps as is done in OP-GaAs methods. Instead, the nonlinear susceptibility is periodically suppressed as in the PSN method, except that no material regrowth is required. Manipulation of the nonlinearity is achieved by using a suite of post-growth processes known as quantum well intermixing (QWI) [76]. For this technique to work, the waveguide core must be composed of a multiple quantum well (MQW) structure that is designed to provide an optical resonance just above the photon energy of the shortest wavelength in the process (the second harmonic in SHG, or the pump wavelength in DFG), such that the nonlinear susceptibility is large while the optical absorption remains low. Using standard lithography, the quantum wells are exposed to a post-growth process that causes their well and barrier layers to blend together as is shown in Fig. 5. This intermixing process alters the composition and dimensions of the quantum wells causing a commensurate increase in the band gap energy. As a result, the resonance features of the $\chi^{(2)}$ dispersion are shifted to high energies, effectively lowering the nonlinearity at the operating wavelengths in the intermixed regions. Since the intermixing effect can be patterned [77], $\chi^{(2)}$ can be modulated in the waveguide core, yielding as-grown and intermixed material sections, as is shown in Fig. 6.

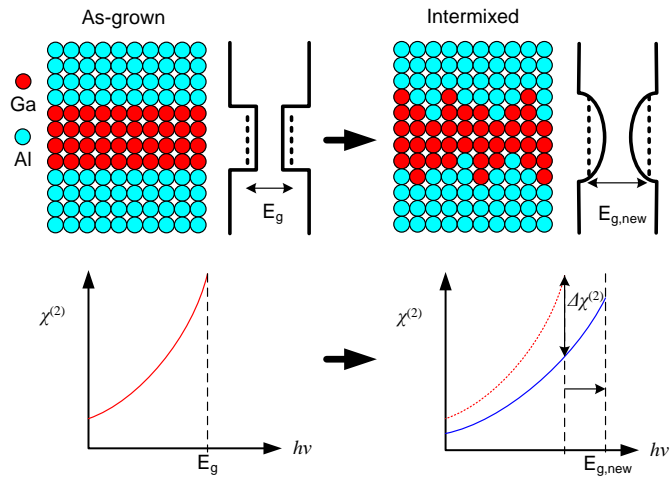

Fig. 5. (Color online) Quantum well intermixing process and the effect on the nonlinear susceptibility. Intermixing increases the band gap energy to $E_{g,new}$ causing a shift in the dispersion of $\chi^{(2)}$ and reducing the magnitude by $\Delta\chi^{(2)}$ at photon energies near the original band gap $E_g$.

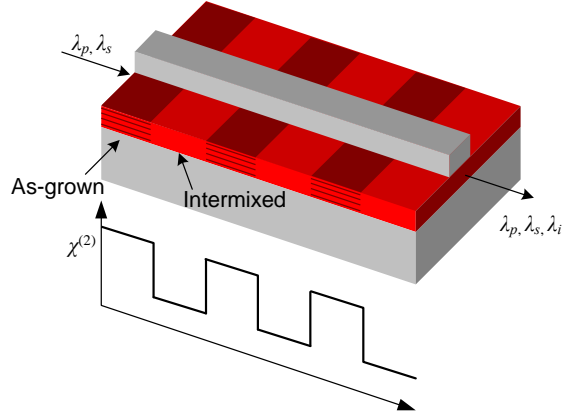

Fig. 6. (Color online) A domain-disordered quasi-phase matching waveguide. As-grown regions retain a large nonlinearity while intermixed regions have a suppressed nonlinearity. Image source: [78].

In a QPM structure, the phase matching condition is satisfied when

$$k_1 - k_2 - k_3 = 2\pi m/\Lambda , \qquad (1)$$

where $k_1$, $k_2$, and $k_3$ are the propagation constants of the three interacting waves, $\Lambda$ is the period of the QPM grating, and $m$ is the order of the grating. In the case of SHG, the phase matching condition becomes

$$k_{2\omega} - k_\omega = 2\pi m/\Lambda , \qquad (2)$$

where $k_\omega$ and $k_{2\omega}$ are the propagation constants of the fundamental and second harmonic waves, respectively. Unlike FBPM and MPM techniques where the phase matching wavelength is generally set by the waveguide heterostructure, the phase matching wavelength in QPM is set during lithography by adjusting the grating period, thus allowing an extra degree of freedom to tailor the operating wavelengths. It permits the possibility of incorporating multiple devices with different phase matching periods onto the same wafer to cover multiple operating wavelengths, and it also allows chirping within an individual grating to increase the phase matching bandwidth. However, the flexibility and lower cost of the QPM technique comes at the expense of a lower potential efficiency. Under ideal conditions, the periodic modulation of $\chi^{(2)}$ leads to a lower effective nonlinearity which is calculated as

$$\chi^{(2)}_{eff} = \Delta\chi^{(2)} \sin(\varsigma\pi)/\pi , \qquad (3)$$

where $\varsigma$ is the grating duty cycle and $\Delta\chi^{(2)}$ is the difference in the value of $\chi^{(2)}$ between the as-grown and intermixed domains. While this effective nonlinearity is at best 60% lower than in FBPM and MPM, in which $\chi^{(2)}_{eff}$ is the material full valued $\chi^{(2)}$, the lower linear loss and large mode overlap in DD-QPM waveguide structures make up for this loss in efficiency. Modulation of $\chi^{(2)}$ in AlGaAs MQW and superlattices is theoretically between 50–100 pm/V, which is comparable to that available in periodically poled lithium niobate (PPLN) [79].

Work on the DD-QPM technique has proceeded over the last number of years with steady improvements to the fabrication processes and waveguide structures leading to increases in device efficiency. Early research used asymmetric quantum well structures, impurity-free vacancy disordering (IFVD) QWI techniques [80]. From there, DD-QPM has evolved to incorporate a symmetric AlGaAs/GaAs superlattice as the waveguide core to increase the maximum amount of band gap and $\chi^{(2)}$ shift [81]. However, the IFVD process is limited in resolution due to lateral spreading of the surface defects induced to promote disorder as shown in Fig. 7(a), and hence restricted devices to high-order gratings that reduced the potential efficiency. Since then, a transition was made to ion-implantation disordering (IID) QWI, shown in Fig. 7(b), in order to increase the resolution of the intermixing process. MicroRaman spectroscopy has confirmed that first-order gratings can be formed by this technique [82]. As a result, higher SHG conversion efficiencies were recorded [83,84]. We have continued the work on DD-QPM and have achieved even greater performance, which will be discussed in Sec.2.2.

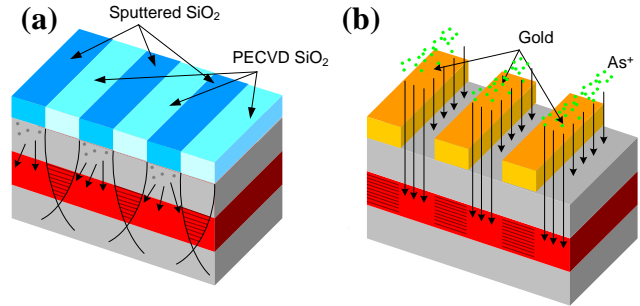

Fig. 7. (Color online) DD-QPM gratings formed using (a) impurity free vacancy disordering via the sputtered silica cap method, and (b) ion-implantation disordering.

### 1.3.3. Microring resonators

A completely different structure that can be exploited for enhancing the optical frequency conversion, but is not limited to this purpose, is the microring resonator. Optical microcavities are in general exploited for many different applications, ranging from lasing sources on a chip (see e.g. [85] and references therein) to frequency comb generation [86,87]. Microcavities have recently attracted a great deal of attention for studying nonlinear wavelength conversion, as they permit to lower thresholds and increase effective conversion efficiencies, thereby allowing CW nonlinear optics. For a comprehensive review on optical microcavities and their applications see e.g. [85]. In this paper we focus on the possibility of exploiting planar integrated microring resonators to lower the power requirements for nonlinear (CW) interactions. In order to describe the efficiency enhancement induced by the resonant structure, we will briefly describe the main properties of FWM in this geometry.

For planar resonant structures (e.g. disks and rings), the electric field of the angular modes are characterized by $E(r,\theta,z,t) \sim A(r,z)exp(-i\omega t + im\theta)$, where $A$ is the spatial distribution of the modes in cylindrical coordinates *(r,θ,z)*, and *ω* is a resonant frequency with integer number *m* [88-90]. By applying Maxwell's equation in the presence of a third-order electric field dependence of the material electric permittivity, one can show that the resulting angular momentum conservation for a degenerate FWM process takes on the following selection rule: *$2m_p = m_s + m_i$* [88-90]. For large resonators with negligible radiation losses, *mθ* is approximately equal to the traditional propagation constant in straight waveguides, *βL*, where *L* is the circumference of the ring/disk resonator. With this approximation, the more commonly reported form of the material phase matching is obtained: *$2\beta_p = \beta_s + \beta_i$*.

Whereas phase matching can be challenging to obtain in straight waveguides, the selection rule (angular momentum conservation) in resonator structures is automatically satisfied for symmetrically distributed idler and signal resonances around the pump resonance (see Fig. 8):

$$m_i - k = m_p = m_s + k, \qquad (4)$$

for some integer *k*.

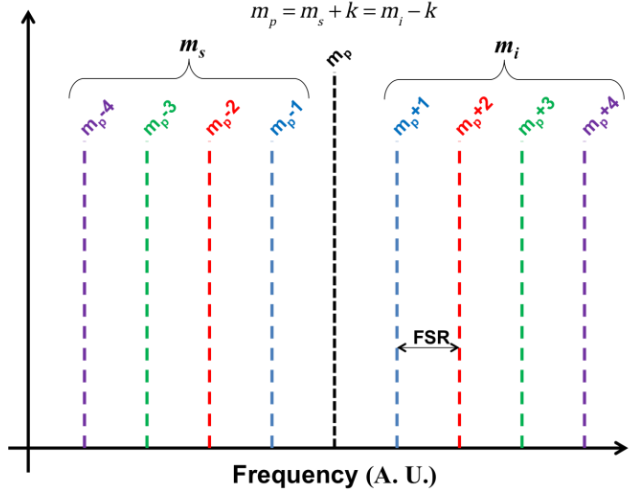

Fig. 8. (Color online) Selection rule for the conservation of angular momentum. The dashed lines represent resonances, also identified by an integer number, whereas the color pairs represent resonances that pair with the pump resonance (black) and that automatically conserve angular momentum.

However, these conditions do not generally guarantee parametric growth. Indeed, in the presence of dispersion, the resonant frequencies are not equally separated by a fixed free spectral range. Consequently, the generated new idler frequency $\omega_i = 2\omega_p - \omega_s$ does not necessarily correspond to a resonance of the structure. This creates a frequency mismatch (analogous to photon energy conservation) of $\Delta\omega = \omega_i - \omega_{ir}$, where $\omega_{ir}$ is the closest cavity resonance to $\omega_i$ [88]. Moreover, contrary to second-order nonlinear optical frequency conversion processes, such as SHG, the requirement of having the material phase matching term (or frequency mismatch for resonators) equal to 0 is not sufficient or always correct for efficient FWM. Indeed, the non-negligible self-action effect of SPM, and the intra-beam interaction of XPM add an intensity dependent term to the net overall phase matching term (or equivalently, the propagation constant of the nonlinear modes is not the same as for the linear modes). This can be intuitively understood from the fact that the Kerr nonlinearity modifies the overall material index, causing a frequency shift in the resonant frequencies that depends on power. Net parametric gain can nevertheless be obtained whenever

$$-4\gamma P_p c/n_g < \Delta\omega < 0 \,, \tag{5}$$

where $\gamma$ is the nonlinear parameter, $P_p$ is the pump power (circulating in the resonator), $c$ is the speed of light in vacuum, and $n_g$ is the group index at $\omega_i$ [15,88], with the maximum gain occurring for $\Delta\omega = -2\gamma P_p c/n_g$. Satisfying this frequency mismatch, along with obeying the selection rule (from angular momentum considerations), is necessary for efficient parametric conversion, although other factors can also limit the process (e.g. polarization, tensor strength and symmetries).

The evolution equations for degenerate FWM in a microring resonator can be obtained using temporal couple mode theory [90,91], or using steady state equations of propagation (when the bending of the resonance can be neglected). In the low-conversion efficiency approximation, assuming negligible frequency mismatch, the net parametric frequency conversion is [92,93]:

$$P_i \approx \gamma^2 P_p^2 P_s L_{\text{eff}}^2 F_E^8 \,, \tag{6}$$

where $P_i$ is the output (coupled out of the resonator) idler power, $P_p$ and $P_s$ are the input pump and signal powers (before being coupled to the resonator), respectively, $L_{\text{eff}}$ is the effective length:

$$L_{\text{eff}} = L\exp(-\alpha L/2)\left(\frac{1-\exp(-\alpha L)}{\alpha L}\right), \tag{7}$$

$\alpha$ is the linear loss coefficient, and $L$ the effective circumference of the resonator. Note that Eq. (5) also assumes that the frequency separation between the pump, signal and idler beams is relatively small such that the spatial modes are geometrically similar. The optical frequency generation process thus depends quadratically on the product of the pump power, the nonlinear parameter, and the effective length, and linearly on the signal power. The term $F_E$ is the field enhancement factor [50,93], and represents the overall contribution of the resonant structure ($F_E$ = 1 for a straight waveguide, where $L$ would be the waveguide length); in the absence of losses for a 4 port ring (as illustrated in Fig. 3), $F_E^2 \approx 2Q \cdot FSR/\omega$, where $FSR$ is the free spectral range and $Q$ is the quality factor. As the conversion efficiency scales with $Q^4$, having large $Q$ factors is extremely beneficial for FWM applications. Unfortunately, a maximum permissible $Q$ exists, at the so-called critical coupling regime where the coupling $Q$ equals the loss $Q$ [94]. This maximum $Q$ is approximately $n\omega/2\alpha c$, and thus it is fundamental to have low losses for high $Q$ factors, and consequently low-power FWM.

The low-conversion efficiency relation, Eq. (6), is useful to understand the physics affecting the growth of the converted signal, but past ~10% conversion, pump depletion becomes significant and numerical means must be used to estimate the net conversion efficiency. However, the upper limit of conversion (i.e. maximal converted power) can be obtained from the Manley-Rowe relations [14,90]. Conceptually, maximum conversion occurs when all pump photons are depleted. The FWM process generates both idler and signal photons, and thus the maximum conversion efficiency is limited to:

$$\eta \equiv \frac{P_i}{P_p} = \frac{\omega_i}{2\omega_p}. \tag{8}$$

When the converted idler frequency is near to the pump frequency, one finds that the maximum conversion efficiency is approximately 50%. In Sec. 3, we report our recent results on optical frequency generation by means of Hydex® microring resonators, exploiting both seeded and spontaneous FWM.

## 2. Optical frequency conversion in AlGaAs waveguides

In this section we present an overview of our recent achievements in optical frequency conversion exploiting second- and third-order nonlinear interactions in AlGaAs strip-loaded waveguides. In particular, we discuss wavelength conversion by SPM, XPM and FWM (Sec. 2.1), as well as by SHG and DFG (Sec. 2.2).

### *2.1. SPM, XPM and FWM in AlGaAs strip-loaded waveguides*

As stated in the introduction, wavelength conversion can be implemented by way of third-order nonlinear effects, such as XPM and FWM, which are the nonlinear optical processes relying on the real part of $\chi^{(3)}$. Together with a high value of $n_2$, a nonlinear material should also exhibit low linear and nonlinear propagation losses in the telecom spectral range (1400 - 1600 nm). The performance of a nonlinear material can be evaluated in terms of the figure of merit (FOM) $T$ [26,36], defined as

$$\frac{1}{T} = \frac{n_2}{\lambda_0 \alpha_2},\qquad(9)$$

where $\lambda_0$ is the free-space wavelength, and $\alpha_2$ is the two-photon absorption (TPA) coefficient resulting from the third-order nonlinearity. For a nonlinear material to be efficient, the condition $T < 1$ must be satisfied [36]. It is worth noting that several FOMs have been defined in literature for characterizing the different nonlinear effects like, *e.g.*, three-photon absorption [36,95].

$Al_{0.18}Ga_{0.82}As$ has the potential for being used in all-optical networks due to its large Kerr coefficient ($n_2 \approx 1.55 \times 10^{-13}$ cm$^2$/W) and small TPA coefficient ($\alpha_2 \approx 0.05$ cm/GW at 1550nm) [36], resulting in a nonlinear figure of merit $T < 0.6$ [26]. In fact, an efficient SPM [35,37,38] and XPM [37] have been demonstrated earlier in AlGaAs ridge waveguides [35] and photonic wires [38] despite the high propagation loss in the latter case. This motivated us to take a step further in optimizing the performance of AlGaAs, making use of its flexibility and tailorability, in order to achieve even more efficient nonlinear interactions.

In our recent studies [47,96], we designed and characterized the nonlinear performance of compact AlGaAs strip-loaded waveguides for efficient wavelength conversion. We chose a strip-loaded waveguide configuration instead of dispersion-engineered submicron waveguides (photonic wires), as it is relatively easy to fabricate by a standard single-step photolithography procedure, and moreover the guided mode in such devices is not sensitive to the fabrication imperfections, as it propagates in the guiding layer underneath the ridge and does not sense the sidewall roughness. We adjusted the AlGaAs wafer composition, as well as the geometrical parameters in our design as to minimize the linear and nonlinear optical absorption losses, while retaining a single-mode operation with minimal effective mode area $A_{\text{eff}}$ to maximize the nonlinear coefficient $\gamma = 2\pi\, n_2/\lambda_0 A_{\text{eff}}$. More specifically, we increased the index contrast between the cladding and the guiding layer compared with previous designs [35,36] to minimize the effective mode area, while choosing the guiding layer composition to be $Al_{0.18}Ga_{0.82}As$, corresponding to a transparency window between the two- and three-photon absorption peaks, within the telecom C-band [36].

Compact strip-loaded AlGaAs waveguides were designed with the wafer composition and waveguide dimensions shown in Fig. 4. We achieved single-mode operation with $A_{\text{eff}}$ =4.3μm$^2$ (TM), $\gamma \approx 13$ m$^{-1}$ W$^{-1}$, measured linear propagation losses of 2–3 dB/cm and a three-photon absorption coefficient $\alpha_3 \approx 0.08 \pm 0.03$ cm$^3$/GW$^2$ in 1.0–2.7-cm-long devices [47]. We did not observe any two-photon absorption in our devices.

Using a tunable optical parametric oscillator (OPO), generating 2-ps FWHM pulses with an average power of 300 mW (peak power of 154W), we experimentally measured the SPM in our waveguides (see Fig. 9). We recorded a nonlinear phase shift of up to 6π, for the TM mode, which is the largest value reported in AlGaAs strip-loaded waveguides to date. This result confirmed that our devices were optimized for highly efficient nonlinear interactions.

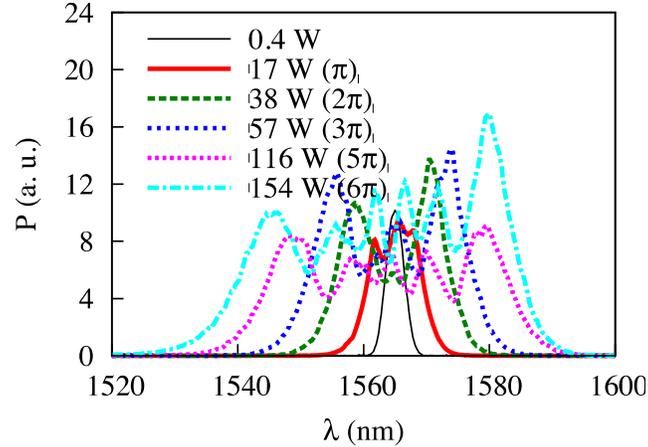

Fig. 9. (Color online) Spectral broadening due to SPM in an AlGaAs strip-loaded waveguide. The legend shows the values of the in-waveguide peak power and the corresponding values of the nonlinear phase shift in brackets. The low-power spectrum exhibiting no broadening is represented with a thin black solid line. Image source: [47].

XPM and FWM was characterized in this device by coupling simultaneously both the OPO pulse and a tunable C-band CW signal laser, amplified by an erbium-doped fiber amplifier (EDFA) to a 2-W level. The high peak power OPO output served as a pump, inducing the nonlinear phase shift in the CW probe (XPM), which resulted in the appearance of side wings around the CW peak [see Fig. 10(a)]. The pump and signal beams also interacted in the sample via FWM, which resulted in the generation of an idler peak on the opposite side of the OPO peak [see Fig.10(a)] at ~1580nm (governed by the FWM energy relation $\omega_i = 2\omega_p - \omega_s$). SHG and DFG were not observed, presumably due to absence of phase matching at the frequencies used.

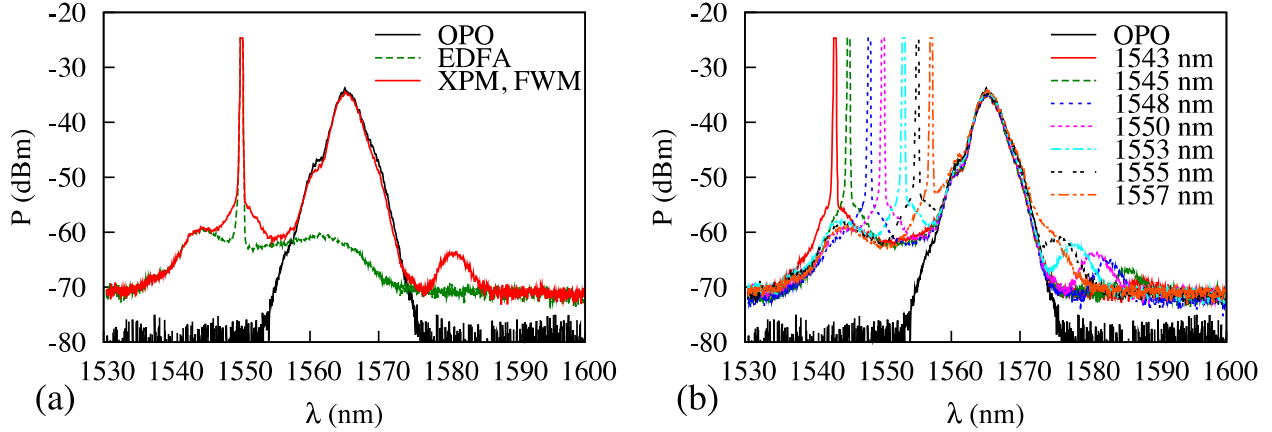

Fig. 10. (Color online) XPM and FWM results. (a) The input signal (amplified CW laser at ~1547nm), green curve, broadens in frequency due to the XPM (red curve) caused by the strong input pump (centered at 1565nm, black curve). FWM is also observed at ~1580nm (red curve). (b) Wavelength tunability of the FWM process, obtained by varying the input signal frequency. Image source: [47].

In Fig.10(b), we show the tunability of the FWM process by varying the wavelength of the CW signal in the range 1543–1557 nm, which was limited not by the bandwidth of the phase matching, which is expected to be wider, but by the gain of the amplifier, that prevented high enough signal powers to take accurate measurements. The maximum tuning range of the CW probe, resulting in the appearance of the generated idler, was measured to be 20 nm [47]. The limiting factors affecting the efficiency of FWM are the material losses and the walk off between the pump and generated idler due to the group velocity dispersion. The walk-off length for the maximum signal-to-idler separation achieved (20 nm) was estimated to be on the order of a few centimeters (from the overall dispersion of our devices), which is on the order of the length of the waveguides in our experiment.

The momentum conservation law that is required for an efficient phase-matched FWM interaction, $\beta_i = 2\beta_p - \beta_s$ is difficult to fulfill because of the material dispersion of AlGaAs: $D_{mat}$ = -1000 ps/nm/km. Note that the waveguide dispersion in our samples is relatively small, changing slightly the total dispersion at 1550 nm to -910 ps/nm/km. Nevertheless, this dispersion-induced phase mismatched was compensated by the induced mismatch from XPM and SPM. We were thus able to achieve a tunability range of up to 14 nm in our experiments with a signal-to-idler conversion efficiency as large as 10dB. Considering other samples, we also achieved a tunability range up to 20nm, with a signal-to-idler conversion efficiency equal to 8db [47]. A maximum separation of 60 nm between the signal and idler wavelengths was observed in our samples, limited by the EDFA tunability range. We thus have achieved good results on wavelength conversion in non-dispersion-managed waveguides. In the next section we discuss the results obtained exploiting phase-matched nonlinear interactions by means of DD-QPM.

## *2.2. SHG and DFG by means of DD-QPM in periodically intermixed waveguides*

Continued interest in using second-order nonlinear effects for wavelength conversion has focused our recent research efforts on improving the efficiency of AlGaAs DD-QPM waveguides. To this end, we have attempted to reduce scattering losses and increase the band gap shift in the superlattice to yield a larger $\Delta\chi^{(2)}$. Additionally, we have made some adjustments to the waveguide structure, including the GaAs/AlGaAs superlattice core layer, to improve the nonlinear behavior. In the SHG experiments performed with the most recent devices, the result has been the production of nearly 10 μW of second-harmonic power using 2-ps optical pulses at wavelengths around 1550 nm in the type-I interaction (TE polarized fundamental, TM second-harmonic), representing a nearly six-times increase over previous generation devices [84] and a conversion efficiency improvement of nearly an order of magnitude over the first demonstrations of superlattice core DD-QPM waveguides [97]. With the improvements made, the latest DD-QPM devices were capable of continuous-wave SHG [68] and type-II SHG (mixed TE/TM fundamental, TE second-harmonic) [98] with internally generated second-harmonic powers in excess of 1 µW. Experiments were also carried out to examine the effect of pulse length on the SHG conversion efficiency, and it was found that high-order nonlinear effects such as nonlinear refraction and two-photon absorption reduced substantially the efficiency at high power [99]. In all experiments, phase matching was achieved at different wavelengths over a span of 60 nm using different QPM periods ranging from 3.1 – 3.8 µm as displayed in Fig. 11, which shows the wide flexibility of DD-QPM.

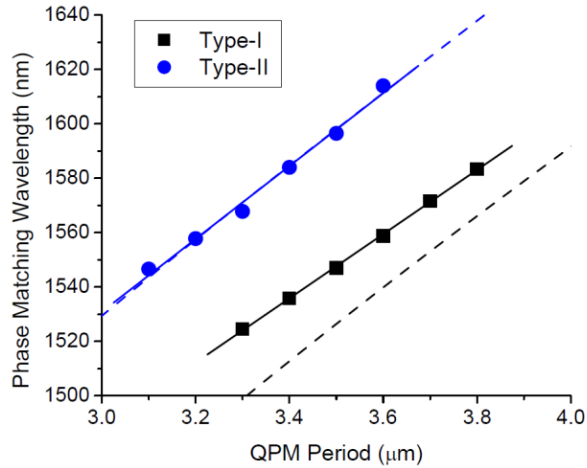

Fig. 11. (Color online) Phase matching wavelengths for SHG in DD-QPM waveguides for different grating periods in both type-I and type-II polarization configurations. Measured data (symbols) and fit trends (solid lines) are shown alongside with the predicted phase matching wavelengths (dashed lines). Image source: [100].

Results from SHG experiments in DD-QPM waveguides can be used to evaluate the strength of the second-order nonlinearity in the AlGaAs/GaAs superlattice core. The normalized conversion efficiency, as defined in Ref. [98], is the usual metric for evaluating performance. DD-QPM waveguides have demonstrated normalized conversion efficiencies of up to 1200 % $W^{-1}cm^{-2}$ for type-I SHG and 350% $W^{-1}cm^{-2}$ for type-II SHG, depending on the phase matching wavelength. After accounting for the impact of linear losses on the efficiency [101] and scaling the values according to the pulse lengths used (2-ps, in this case), the value of $\chi^{(2)}_{eff}$ and $\Delta\chi^{(2)}$ can be extracted. Figure 12 shows the change in the susceptibility tensor elements $\chi^{(2)}_{zxy}$ and $\chi^{(2)}_{xyz}$, which are associated with the type-I and type-II interactions, respectively, for the AlGaAs/GaAs superlattice core. The modulation is as high as 26 pm/V, which is about half the expected amount for this superlattice structure and below what is achievable in PPLN. This can be explained as the result of only partially intermixing the superlattice resulting in a band gap differential between as-grown and intermixed areas of only 27% of the potential maximum. Additional optimization of the QWI process will be necessary to yield the needed band gap shift and $\Delta\chi^{(2)}$ for competitive efficiency.

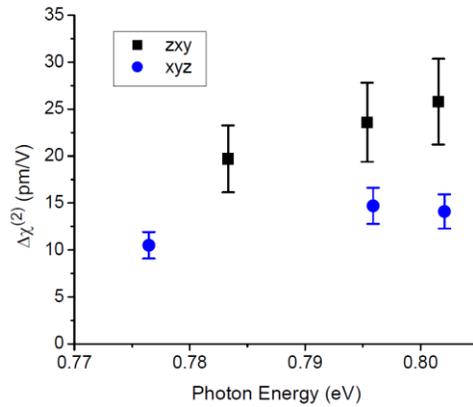

Fig. 12. (Color online) Modulation in $\chi^{(2)}$ between as-grown and intermixed AlGaAs/GaAs superlattices.

Most recently, we have made the first demonstration of DFG in DD-QPM waveguides between wavelengths in the C-, L-, and U-band [100]. We were able to show both type-I and type-II phase matching based on the polarization configurations of the signal, idler, and pump beams. As shown in Fig. 13, the wavelengths were converted over a span of up to 100 nm with a single unchirped DD-QPM grating. Such a large bandwidth would make it suitable for wavelength conversion in DWDM systems. Conversion efficiencies were comparatively low at around -65 dB. However, considering that the input beams were continuous waves, this is a rather good result for a first attempt. Further improvements to lower the losses and to improve the physics of the QWI process will lead to more practical efficiency levels.

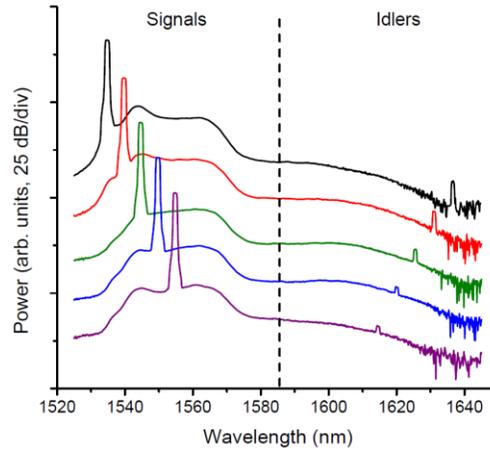

Fig. 13. (Color online) Output spectra for type-I phase matched DFG in a DD-QPM waveguide with different input wavelengths. Reproduced from [100].

The goal of using DD-QPM in AlGaAs/GaAs superlattices is to build a platform for complete, monolithically integrated optical frequency conversion devices. As depicted in the example of Fig. 14, such devices would include a pump laser to drive the nonlinear conversion process, dichroic couplers for combining and splitting the pump and signal wavelengths, and passive waveguides for routing the signals on and off the chip. All of these components can be built in the same AlGaAs material system without the need for hybrid material integration and without the need for etch-and-regrowth processes. In addition to altering the nonlinear properties of the superlattice core, band gap shifting by QWI can be used to mix active and passive areas onto the same chip. Superlattice-core lasers with a similar layer configuration as the DD-QPM devices have been already demonstrated [102]. Dichroic couplers based on multi-mode interference (MMI) structures have also been fabricated and show sufficient separation of the pump and signal wavelengths. Though some challenges remain, these building blocks are suitable for integration with each other for a true all-optical PIC solution to wavelength conversion.

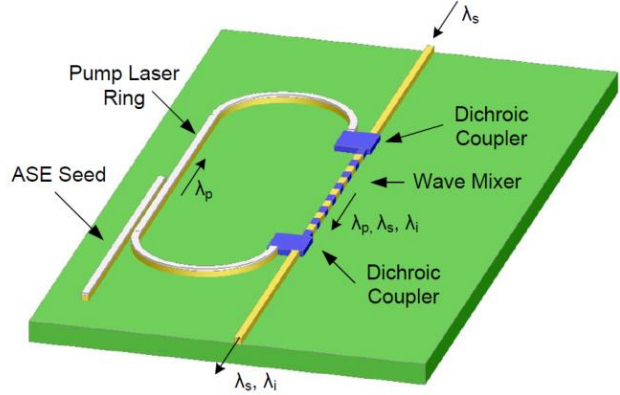

Fig. 14. (Color online) Schematic of a fully integrated all-optical wavelength converter based on a wave mixer in the form of a DD-QPM waveguide.

## 3. Optical frequency conversion in Hydex®

In this section we overview our recent achievements in frequency conversion exploiting FWM in Hydex® microring resonators. We discuss both the case in which the seed is externally provided (Sec. 3.1), as well as the case in which vacuum fluctuations act as a seed for a spontaneous FWM process (Sec. 3.2).

### *3.1. CW Four Wave Mixing*

Exploiting the ultra-low losses (0.06dB/cm) of the Hydex® platform, a 4-port Hydex® microring resonator with a radius of 135μm was fabricated as shown in Fig. 3. Whereas the maximum theoretical $Q$ is approximately $2.5 \times 10^6$ for this platform, the experimentally measured $Q$ for our design was approximately half this value ($\approx 1.2 \times 10^6$) with a corresponding resonance linewidth of 1.3pm (160MHz) at 1550nm, and a free spectral range of 200GHz. The dispersion in this platform was evaluated by measuring the frequency dependence of the free-spectral range, from which we numerically estimated the maximum allowable FWM bandwidth [50]. Figure 15 portrays the frequency mismatch factor ($\Delta\omega$) as a function of the pump frequency and the signal frequency (for low input pump and signal powers). The colored region indicates the condition when $\Delta\omega$ < 80MHz, which is half of the resonance full width at half maximum (FWHM) linewidth, and thus where FWM is frequency matched with the ring resonances. In contrast, the black region indicates frequency combinations where FWM is not realizable. As can be seen from the figure, the zero dispersion point is near 1560 nm in this structure (vertical polarized mode), whereas the curvature in the plot is a result of fourth order dispersion terms that were taken into account for the modeling [50]. A FWM bandwidth broader than 160nm is numerically extrapolated from the experimental data for an operative pump frequency near the zero-dispersion point (see [50] for more details). This remarkable result is a consequence of the low dispersive properties of our platform. Note that whereas the power dependence of the total mismatch is not represented in Fig. 15 (obtained for an arbitrary low power), the net effect of the SPM/XPM induced frequency mismatch will simply be to offset the figure slightly since this mismatch is (approximately) frequency independent.

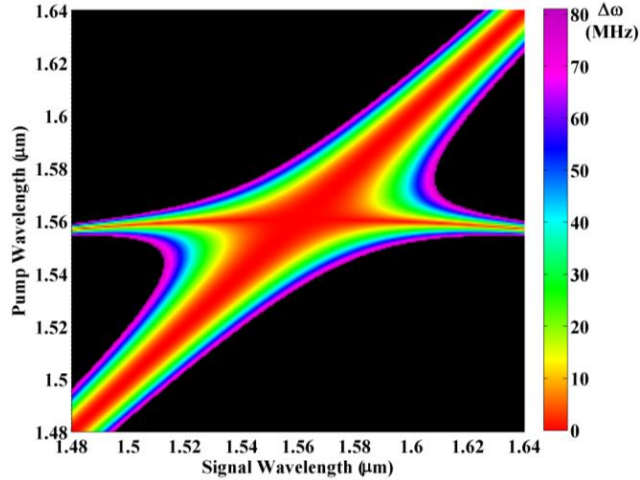

Fig. 15. (Color online) Frequency mismatch in the Hydex® high $Q$ ring resonator. The color intensity plot represents the net frequency mismatch as a function of the excited pump and signal frequencies (resonances). The region where the frequency mismatch is too large to observe tangible FWM is represented in black. Note that we interpolated the data between resonances for aesthetics and to add clarity in reading the figure.

FWM was investigated by using two low power CW lasers. Experimentally, each laser was polarized and frequency aligned to a separate cavity resonance; the pump laser was used to excite the Input port (see Fig. 3), whereas the signal laser was used to excite the Add port. The output power and spectrum (using an optical spectrum analyzer) were collected at the Drop port in these experiments, although at resonance the pump, signal, and converted idler beams exit from both the Drop and Through ports: see [50] for further details on the experimental setup. Figure 16 shows the results of two separate experiments. In the first experiment the pump and the signal were placed at adjacent resonances ($\lambda_p$ = 1551.36 nm, $\lambda_s$ = 1552.99 nm) with laser powers of 8.8 mW and 1.25 mW, respectively (note that the coupling efficiency into the bus waveguides is approximately 70%). Frequency conversion via FWM produced a first idler ($\lambda_i$ =1549.75 nm) with an overall conversion efficiency $P_i/P_s$ = -36 dB, corresponding to an idler of approximately 130 nW (out of the device). FWM using non-amplified low-power CW light is rather impressive, as nonlinear optics is typically coined with high power laser sources. Our results stem directly from the large field enhancement factor obtained, *i.e.* $F_E$ = 17.9. From Eq. (6), this corresponds to an efficiency increase of ~1 x $10^{10}$. Obtaining such a high contribution, or conversion efficiency, in a longer straight waveguide with these low losses is not possible due to the saturation of the effective interaction length.

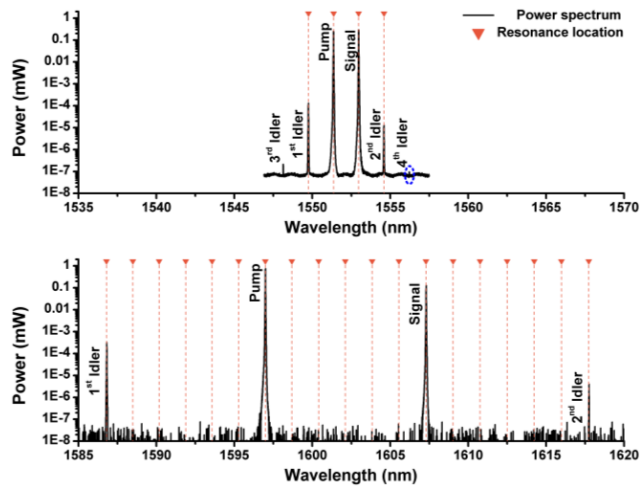

Fig. 16. (Color online) FWM experimental results for two different excitation conditions; dashed lines correspond to resonant frequencies. (upper) The pump and signal lasers are tuned to adjacent resonances and generate a series of idlers (through cascaded FWM) that are neighboring resonances. (lower) The pump and signal lasers are tuned to resonances that are 7 resonances apart (1.2 THz), resulting in idlers that are also separated by this frequency spacing. Image source [50].

The efficiency of the process was also sufficient to see a cascaded process of FWM whereby the signal, pump (which in this degenerate FWM process is defined to be the source of two of the three input photons needed for the FWM interaction), and idler formally exchange roles to produce a new idler (see Fig. 16). This cascaded process produced a total of 4 generated detectable idlers [50]. The experiment was also repeated for signal and pump resonances 6 FSRs apart (1.2 THz) ($\lambda_p$ = 1596.99 nm, $\lambda_s$ = 1607.31 nm), obtaining the same efficiency within the experimental error. By tuning the signal off resonance, the idler was deduced to be exactly on resonance (within experimental accuracy) when the signal and pump were themselves on resonance, validating the low dispersion results predicted from Fig. 15. Moreover, the quadratic power dependence of the pump power on the generated idler power, as well as the linear dependence of the signal power on the idler power, as dictated by Eq. (6), were verified experimentally [93].

This [50] and other results reported by our group [93] represents the first CW nonlinear optics experiment in a silica-glass based integrated platform. Our promising results are a direct consequence of the material platform and low losses permissible. Similar results have been recently proved to be also possible in silicon nitride [61,103], a high-index glass platform characyerized by extremely low losses. Both Hydex® and silicon nitride offers the possibility of fabricating high $Q$ structures, with absence of multiphoton absorption, allowing even higher conversion efficiencies than reported here through the use of higher input pump powers. For comparison, CW FWM was also demonstrated in silicon-on-insulator microring resonators with only 5mW of coupled pump power [93], obtaining a conversion efficiency of -25.4 dB.

Unfortunately, the FWM conversion efficiency was found to saturate at higher pump powers due to two-photon absorption and to the generated free carriers [93], placing overall limitations on the nonlinear gain, which presumably is the reason why a silicon OPO has yet to be reported.

Many FWM experiments have equally been performed with pulses, where waveguide devices are preferred over cavities due to the limited bandwidth of typical resonances. Multiple applications in a host of optical platforms have been reported, including all-optical regeneration and demultiplexing in chalcogenides [104,105], silicon [52], GaAs [47,96], silica photonic crystal fibers [106], as well as in Hydex® [55,107]. A direct application of integrated FWM is in telecommunication systems, where signals may need to be remapped to different frequencies (*i.e.*, in the case of wavelength contention) for nonstop service to the consumer [108].

## *3.2. Optical hyper parametric oscillator*

For certain applications, the requirement of two lasers for nonlinear frequency mixing can be problematic, cumbersome and costly. For instance, attosecond physics [109], molecular fingerprinting [110], broadband sensing and spectroscopy [111], and optical interconnects would make great use of a compact multiple wavelength source, where a single (on-chip or off-chip) CW laser source could generate a cascade of CW frequency components to be either modulated, sampled or used for synchronization on-chip. Such a source can be realized from a high Q resonator with the appropriate nonlinear gain.

Figure 17 examines the typical power dependence of the FWM process as a function of the pump and signal power in the absence of the SPM/XPM terms. Interestingly, one can observe that for an arbitrarily low signal power, quantum limited conversion efficiency remains possible [90]. We have to consider that the SPM/XPM terms cause the resonances to shift, thereby preventing the results of Fig. 17 (at least on first observation). However, if one considers that the resonance have a finite linewidth, and that the pump laser can be tuned continuously (which is possible experimentally) to account for the power dependent shift in the resonances, the conceptual results of Fig. 17 remain valid [90]. In the limit of no signal power, the quantum vacuum replaces the signal source, and spontaneous FWM can be obtained. The dispersion and power dependence of the frequency mismatch relation presented in Sec. 1.3.3 and in Eq. (5) are still maintained. Two photons from a single laser source can initiate the creation of a signal and idler pair, with vacuum acting to spontaneously seed the process. In a low-loss resonator with sufficient pump power, the nonlinear gain of the generated broadband idler signal can surpass the overall net losses (coupling to the ports, bending and propagation loss), such that oscillations can occur. The process is similar to a laser, albeit mediated by virtual states instead of the real quantum states of the system [14]. At the threshold power for oscillation, the gain peaks for a narrowband spectral region around a signal and idler pair. Supplying more pump power causes an exponential increase in the generated idler and signal, allowing for the cavity to self-seed and oscillate. This type of resonant oscillation behavior is commonly referred to as an optical parametric oscillator when the interaction is a result from difference frequency generation in a $\chi^{(2)}$ medium [14]. For a third order process, such as spontaneous FWM, the device is called an optical hyper-parametric oscillator [113].

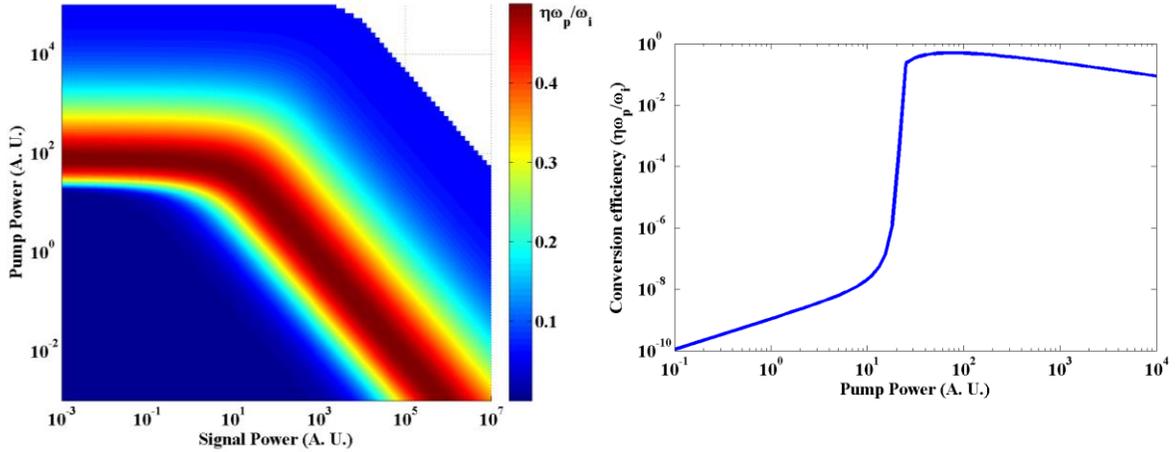

Fig. 17. (Color online) (left) Conversion efficiency of FWM, neglecting the effects of SPM/XPM, in a micro-ring resonator as a function of the input pump and idler powers. As the signal power is reduced to arbitrary low values, maximal frequency conversion remains viable at a fixed pump power (~100 in the figure). (right) Plot of the efficiency of FWM in a microring resonator for a low signal power of 1pW (blue line), showing a threshold power for oscillation (~20 in the figure).

The threshold power required for oscillation, and the corresponding signal-idler pair that oscillates, can be estimated using couple mode theory [90], or directly by equating the FWM gain to the losses within the cavity for one complete cycle [114]. The threshold power was estimated to be ~50 mW (coupled) for the Hydex® high $Q$ resonator when the pump was placed at a ring resonance of 1544.15 nm, and the signal and idler pairs were expected to oscillate roughly 50 nm from the pump [114]. Experimentally, oscillation was observed by using a single CW fiber laser, amplified via an erbium doped fiber amplifier, coupled to the Input port of the resonator (see Fig. 3). The significantly higher optical power used in this experiment (roughly ten times higher than the value used in the seeded degenerate FWM experiments) causes considerable amount of cavity heating, changing the resonances, in turn leading to hysteresis and some instabilities [115]. To compensate for these effects, the resonator was thermally isolated on a Peltier cell to control and maintain its temperature, and a thermal locking procedure was used to slowly bring the pump laser into resonance [114]. Incidentally, this continuous tuning of the pump frequency compensates experimentally for the frequency mismatch introduced by SPM and XPM. The output of the ring resonator was collected at the Drop port using a power meter and optical spectrum analyzer.

Experimentally, the threshold power was determined to be 54 mW, with the corresponding signal-idler pair separated by 52 FSR ($\lambda_s$ = 1596.98 nm, $\lambda_i$ =1494.70 nm), as shown in Fig. 18. The nonlinear conversion from the pump to the new frequencies is extremely critical at pump powers near the threshold power. Above threshold, an increase in the pump power causes the idler (and signal) to grow linearly, see Fig. 19, and also marks the onset of cascaded FWM. The linear growth plateaus at higher input powers as the quantum limited conversion efficiency (with losses) is approached, and as cascaded FWM becomes increasingly important. The differential slope efficiency (per port and for a single line) directly above threshold was measured to be $P_i/P_p$ = 7.4%, whereas the total maximum conversion was found to be 9mW in all oscillating modes [114]. Furthermore, the wide frequency spacing amongst the oscillating modes represents a bandwidth of more than 6THz. It was also shown that pumping different resonances can lead to significant changes in the oscillation conditions and threshold powers, presumably due to the proximity of the zero dispersion point.

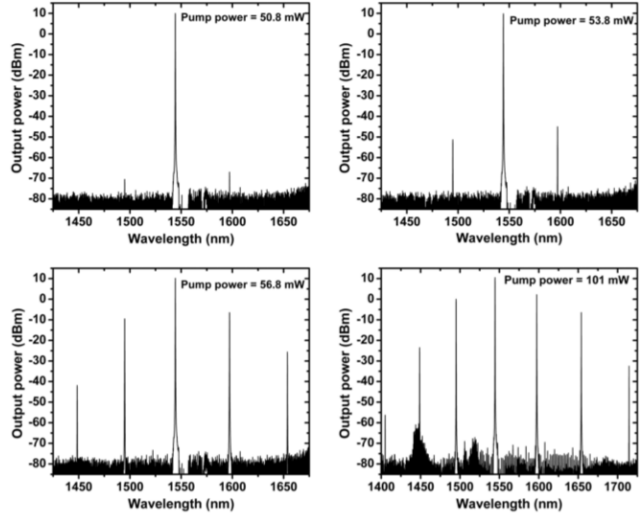

Fig. 18. Experimentally obtained parametric oscillation in the Hydex® microring resonator as a function of the pump ($\lambda_p$ = 1544.15 nm) power. At threshold ($P_p$~54 mW) we found that a pair of idlers ($\lambda_s$ = 1596.98 nm, $\lambda_i$ =1494.70 nm) that grow out of noise with an exponential dependence on the pump power is generated. Cascaded FWM can also be observed above threshold, creating another pair of idlers (~1448 nm and ~1654 nm).

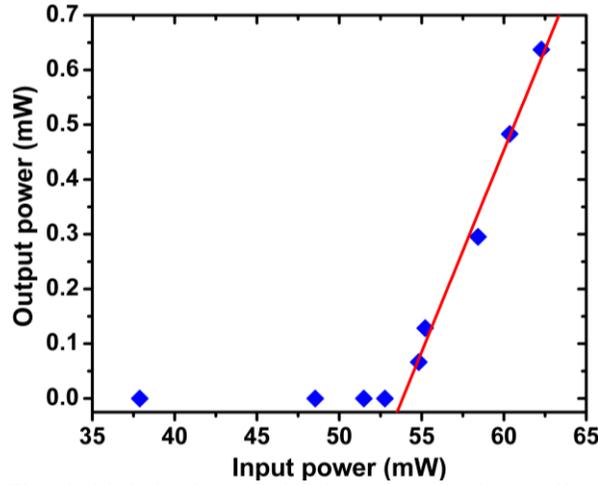

Fig. 19. (Color online) Threshold behavior of the hyperparametric oscillator. Threshold is found at approximately 54 mW of pump power, and the differential slope efficiency (for the Drop channel exclusively and for a single line at 1596.98nm) was determined to be 7.4%.

Together with recent results in similar silicon nitride based microring resonators [116], this is the first demonstration of an integrated optical parametric oscillator, whilst other demonstrations in non-integrated platforms have also been shown, such as in $CaF_2$ [113] and silica microtoroids [89], and in fibers and photonic crystal fibers [117,118]. The smooth and robust fabrication process of these non-integrated platform allow for even lower losses than those obtained with Hydex®, and consequently higher $Q$ factors and higher efficiencies. However, on-chip CMOS compatible systems are generally desired for numerous applications that prioritize integrating numerous devices/operations on a single low-cost multifunctional chip. Direct applications include the potential to replace multiple lasers in optical WDM systems with a single laser source. As recently shown by Intel [119], this CMOS compatible technology would be extremely cost and space effective in replacing multiple hybrid silicon lasers via a high $Q$ microring resonator.

## 4. Conclusions

In this paper we have discussed our recent progress in frequency conversion using AlGaAs strip-loaded waveguides and Hydex® microring resonators. Both material platforms have exceptionally good optical properties, such as negligible linear and nonlinear optical absorption, low propagation losses (Hydex®), a high value of the Kerr nonlinearity, and significant flexibility and tailorability (AlGaAs). Among our recent achievements are the demonstration of broadband self-phase modulation and efficient cross-phase modulation and four-wave mixing in AlGaAs strip-loaded waveguides. We have also successfully realized phase-matched second-order nonlinear interactions, such as difference-frequency and second-harmonic generation, in domain-disordered quasi-phase-matched AlGaAs strip-loaded waveguides. Our recent achievements in high-index doped-glass include the demonstration of CW four-wave mixing and hyper-parametric oscillation in microring resonators. It is worth mentioning that we also observed efficient SHG in AlGaAs photonics wires [72], as well as efficient self-phase modulation [49], supercontinuum generation [59], and FWM in Hydex® spiral waveguides, which we did not discuss in this review due to space limitations.

In conclusion, we strongly believe that the results presented confirm that these two material platforms have much to offer for future all-optical networks "on a chip". As such, the increasingly numerous attempts to realize highly-efficient integrated nonlinear optical devices in AlGaAs and Hydex® are of significant importance.

## 5. Acknowledgments

We acknowledge B. E. Little and S. T. Chu for providing us with the Hydex® samples and CMC Mycrosystems for the wafers growth. We also acknowledge the efforts of A. Helmy of the University of Toronto, of B. M. Holmes, D. C. Hutchings, M. Sorel and the James Watt Nanofabrication Center of the University of Glasgow, U.K., and of U. Younis of the National University of Science and Technology, Pakistan. The photolithography of the AlGaAs samples was performed in part at the ECTI fabrication facility at the University of Toronto. This work was supported by the Natural Sciences and Engineering Research Council of Canada (NSERC), by the Fonds de Recherche du Québec – Nature et Technologies (FQRNT), by the Canadian Institute for Photonic Innovations (CIPI) and by the Ministère du Développement Économique, Innovation et Exportation (MDEIE).

LC acknowledges the support from the Government of Canada through the PDRF program. The research of KD is supported by Canadian Mitacs Elevate postdoctoral fellowship. The work of SJ Wagner was supported by an NSERC Postgraduate Scholarship. DD and MF wish to acknowledge the NSERC –contracts no. PDF-387430-2010 and PDF-387780-2010, respectively– for supporting their research activities. DJM acknowledges support from the Australian Research Council Centres of Excellence and Discovery Project research programs.

# 6. References

1. P. Franken, A. Hill, C. Peters, and G. Weinreich, "Generation of Optical Harmonics," Phys. Rev. Lett. **7**, 118-119 (1961).

2. J. Armstrong, N. Bloembergen, J. Ducuing, and P. Pershan, "Interactions between Light Waves in a Nonlinear Dielectric," Phys. Rev. **127**, 1918-1939 (1962).

3. N. Bloembergen and P. Pershan, "Light Waves at the Boundary of Nonlinear Media," Phys. Rev. **128**, 606-622 (1962).

4. P. K. Tien, "Integrated optics and new wave phenomena in optical waveguides," Rev. Mod. Phys. **49**, 361-420 (1977).

5. G. I. Stegeman and C. T. Seaton, "Nonlinear integrated optics," J. Appl. Phys. **58**, R57-R78 (1985).

6. W. Sohler, B. Hampel, R. Regener, R. Ricken, H. Suche, and R. Volk, "Integrated optical parametric devices," J. Lightwave Technol. **4**, 772-777 (1986).

7. G. I. Stegeman, E. M. Wright, N. Finlayson, R. Zanoni, and C. T. Seaton, "Third order nonlinear integrated optics," J. Lightwave Technol. **6**, 953-970 (1988).

8. D. B. Anderson and T. J. Boyd, "Wideband $CO_2$ Laser Second Harmonic Generation Phase Matched in GaAs Thin-Film Waveguides," Appl. Phys. Lett. **19**, 266-268 (1971).

9. A simple search on the Scopus database for "frequency conversion" or "wavelength conversion" returns more than 26,000 publications in the last decade (2001-2010).

10. O. Leclerc, B. Lavigne, E. Balmefrezol, P. Brindel, L. Pierre, D. Rouvillain, and F. Seguineau, "Optical Regeneration at 40 Gb/s and Beyond," J. Lightwave Technol. **21**, 2779-2790 (2003).

11. S. J. B. Yoo, "Wavelength conversion technologies for WDM network applications," J. Lightwave Technol. **14**, 955-966 (1996).

12. B. Ramamurthy and B. Mukherjee, "Wavelength conversion in WDM networking," IEEE J. Sel. Areas Commun. **16**, 1061-1073 (1998).

13. S. Venugopal Rao, K. Moutzouris, and M. Ebrahimzadeh, "Nonlinear frequency conversion in semiconductor optical waveguides using birefringent, modal and quasi-phase-matching techniques," J. Opt. A: Pure Appl. Opt. **6**, 569-584 (2004).

14. R. W. Boyd, *Nonlinear Optics*, 3rd ed. (Academic Press, 2008), p. 640.

15. G. Agrawal, *Nonlinear Fiber Optics*, 3rd ed. (Academic Press, 2001), p. 467.

16. G. I. Stegeman, *Nonlinear Guided Wave Optics* (Wiley-Blackwell, 1998), p. 500.

17. F. Shimizu, "Frequency Broadening in Liquids by a Short Light Pulse," Phys. Rev. Lett. **19**, 1097-1100 (1967).

18. F. DeMartini, C. Townes, T. Gustafson, and P. Kelley, "Self-Steepening of Light Pulses," Phys. Rev. **164**, 312-323 (1967).

19. G. Agrawal, *Applications of Nonlinear Fiber Optics*, 2nd ed. (Academic Press, 2008), p. 528.

20. S. Radic, D. J. Moss, and B. J. Eggleton, "Nonlinear optics in communications: From crippling impairment to ultrafast tools," in *Optical Fiber Telecommunications V A: Components and Subsystems*, I. P. Kaminow, T. Li, and A. E. Willner, eds., 5th ed. (Academic Press, 2008), pp. 759-828.

21. A. Alduino and M. Paniccia, "Interconnects: Wiring electronics with light," Nat. Photonics **1**, 153-155 (2007).

22. T. L. Koch and U. Koren, "Semiconductor photonic integrated circuits," IEEE J. Quantum Electron. **27**, 641-653 (1991).

23. B. E. Little and S. T. Chu, "Toward Very Large-Scale Integrated Photonics," Optics and Photonics News **11**, 24 (2000).

24. B. Jalali and S. Fathpour, "Silicon Photonics," J. Lightwave Technol. **24**, 4600-4615 (2006).

25. L. Tsybeskov, D. J. Lockwood, and M. Ichikawa, "Silicon Photonics: CMOS Going Optical," Proc. IEEE **97**, 1161-1165 (2009).

26. G. I. Stegeman, A. Villeneuve, J. Kang, J. S. Aitchison, C. N. Ironside, K. Al-Hemyari, C. C. Yang, C.-H. Lin, H.-H. Lin, G. T. Kennedy, R. S. Grant, and W. Sibbett, "AlGaAs Below Half Bandgap: the Silicon of Nonlinear Optical Materials," Journal of Nonlinear Optical Physics & Materials **3**, 347-371 (1994).

27. V. Ta'eed, N. J. Baker, L. Fu, K. Finsterbusch, M. R. E. Lamont, D. J. Moss, H. C. Nguyen, B. J. Eggleton, D.-Y. Choi, S. Madden, and B. Luther-Davies, "Ultrafast all-optical chalcogenide glass photonic circuits," Opt. Express **15**, 9205-9221 (2007).

28. H. W. M. Salemink, F. Horst, R. Germann, B. J. Offrein, and G. L. Bona, "Silicon-Oxynitride (SiON) for Photonic Integrated Circuits," MRS Proceedings **574**, 255-260 (1999).

29. J. H. Lee, K. Kikuchi, T. Nagashima, T. Hasegawa, S. Ohara, and N. Sugimoto, "All-fiber 80-Gbit/s wavelength converter using 1-m-long Bismuth Oxide-based nonlinear optical fiber with a nonlinearity gamma of 1100 $W^{-1}$ $km^{-1}$," Opt. Express **13**, 3144-3149 (2005).

30. M. A. Foster, A. C. Turner, M. Lipson, and A. L. Gaeta, "Nonlinear optics in photonic nanowires," Opt. Express **16**, 1300-1320 (2008).

31. E. Dulkeith, Y. A. Vlasov, X. Chen, N. C. Panoiu, and R. M. J. Osgood, "Self-phase-modulation in submicron silicon-on-insulator photonic wires," Opt. Express **14**, 5524-5534 (2006).

32. T. K. Liang and H. K. Tsang, "Role of free carriers from two-photon absorption in Raman amplification in silicon-on-insulator waveguides," Appl. Phys. Lett. **84**, 2745-2747 (2004).

33. H. K. Tsang and Y. Liu, "Nonlinear optical properties of silicon waveguides," Semicond. Sci. Technol. **23**, 064007 (2008).

34. Brent Little, "A VLSI Photonics Platform," in *Optical Fiber Communication Conference* (Optical Society of America, 2003), pp. 444-445.

35. A. Villeneuve, J. S. Aitchison, B. Vögele, R. Tapella, J. U. Kang, C. Trevino, and G. I. Stegeman, "Waveguide design for minimum nonlinear effective area and switching energy in AlGaAs at half the bandgap," Electron. Lett **31**, 549-551 (1995).

36. J. S. Aitchison, D. C. Hutchings, J. U. Kang, G. I. Stegeman, and A. Villeneuve, "The nonlinear optical properties of AlGaAs at the half band gap," IEEE J. Quantum Electron. **33**, 341-348 (1997).

37. D. Duchesne, R. Morandotti, G. A. Siviloglou, R. El-Ganainy, G. I. Stegeman, D. N. Christodoulides, D. Modotto, A. Locatelli, C. De Angelis, F. Pozzi, and M. Sorel, "Nonlinear Photonics in AlGaAs Photonics Nanowires: Self Phase and Cross Phase Modulation," in *2007 International Symposium on Signals, Systems and Electronics* (IEEE, 2007), pp. 475-478.

38. G. A. Siviloglou, S. Suntsov, R. El-Ganainy, R. Iwanow, G. I. Stegeman, D. N. Christodoulides, R. Morandotti, D. Modotto, A. Locatelli, C. De Angelis, F. Pozzi, C. R. Stanley, and M. Sorel, "Enhanced third-order nonlinear effects in optical AlGaAs nanowires," Opt. Express **14**, 9377-9384 (2006).

39. M. Choy and R. Byer, "Accurate second-order susceptibility measurements of visible and infrared nonlinear crystals," Phys. Rev. B **14**, 1693-1706 (1976).

40. A. S. Helmy, P. Abolghasem, J. Stewart Aitchison, B. J. Bijlani, J. Han, B. M. Holmes, D. C. Hutchings, U. Younis, and S. J. Wagner, "Recent advances in phase matching of second-order nonlinearities in monolithic semiconductor waveguides," Laser Photonics Rev. **5**, 272-286 (2011).

41. Q. Lin, O. J. Painter, and G. P. Agrawal, "Nonlinear optical phenomena in silicon waveguides: modeling and applications," Opt. Express **15**, 16604-16644 (2007).

42. V. G. Dmitriev, G. G. Gurzadyan, and D. N. Nikogosyan, *Handbook of Nonlinear Optical Crystals*, 3rd ed. (Springer, 1999), p. 431.

43. E. C. Mägi, L. B. Fu, H. C. Nguyen, M. R. Lamont, D. I. Yeom, and B. J. Eggleton, "Enhanced Kerr nonlinearity in sub-wavelength diameter $As_2Se_3$ chalcogenide fiber tapers," Opt. Express **15**, 10324-10329 (2007).

44. A. Zakery and S. R. Elliott, "Optical properties and applications of chalcogenide glasses: a review," J. Non-Cryst. Solids **330**, 1-12 (2003).

45. Y. A. Vlasov and S. J. McNab, "Losses in single-mode silicon-on-insulator strip waveguides and bends," Opt. Express **12**, 1622-1631 (2004).

46. A. C. Turner, M. A. Foster, A. L. Gaeta, and M. Lipson, "Ultra-low power parametric frequency conversion in a silicon microring resonator," Opt. Express **16**, 4881 (2008).

47. K. Dolgaleva, W. C. Ng, L. Qian, and J. S. Aitchison, "Compact highly-nonlinear AlGaAs waveguides for efficient wavelength conversion," Opt. Express **19**, 1496-1498 (2011).

48. L.-W. Luo, G. S. Wiederhecker, J. Cardenas, C. Poitras, and M. Lipson, "High quality factor etchless silicon photonic ring resonators," Opt. Express **19**, 6284-6289 (2011).

49. D. Duchesne, M. Ferrera, L. Razzari, R. Morandotti, B. E. Little, S. T. Chu, and D. J. Moss, "Efficient self-phase modulation in low loss, high index doped silica glass integrated waveguides," Opt. Express **17**, 1865-1870 (2009).

50. M. Ferrera, D. Duchesne, L. Razzari, M. Peccianti, R. Morandotti, P. Cheben, S. Janz, D.-X. Xu, B. E. Little, S. Chu, and D. J. Moss, "Low power four wave mixing in an integrated, micro-ring resonator with Q= 1.2 million," Opt. Express **17**, 14098–14103 (2009).

51. M. A. Foster, A. C. Turner, J. E. Sharping, B. S. Schmidt, M. Lipson, and A. L. Gaeta, "Broad-band optical parametric gain on a silicon photonic chip," Nature **441**, 960-963 (2006).

52. R. Salem, M. A. Foster, A. C. Turner, D. F. Geraghty, M. Lipson, and A. L. Gaeta, "All-optical regeneration on a silicon chip," Opt. Express **15**, 7802-7809 (2007).

53. B. E. Little, S. T. Chu, P. P. Absil, J. V. Hryniewicz, F. G. Johnson, F. Seiferth, D. Gill, V. Van, O. King, and M. Trakalo, "Very High-Order Microring Resonator Filters for WDM Applications," IEEE Photonics Technol. Lett. **16**, 2263-2265 (2004).

54. A. Yalcin, K. C. Popat, J. C. Aldridge, T. A. Desai, J. Hryniewicz, N. Chbouki, B. E. Little, V. Van, D. Gill, M. Anthes-Washburn, M. S. Unlu, and B. B. Goldberg, "Optical sensing of biomolecules using microring resonators," IEEE J. Sel. Top. Quant. **12**, 148-155 (2006).

55. A. Pasquazi, R. Ahmad, M. Rochette, M. Lamont, B. E. Little, S. T. Chu, R. Morandotti, and D. J. Moss, "All-optical wavelength conversion in an integrated ring resonator," Opt. Express **18**, 3858 (2010).

56. M. Ferrera, Y. Park, L. Razzari, B. E. Little, S. T. Chu, R. Morandotti, D. J. Moss, and J. Azaña, "On-chip CMOS-compatible all-optical integrator," Nat. Commun. **1**, 29 (2010).

57. M. Peccianti, M. Ferrera, L. Razzari, R. Morandotti, B. E. Little, S. T. Chu, and D. J. Moss, "Subpicosecond optical pulse compression via an integrated nonlinear chirper," Opt. Express **18**, 7625-7633 (2010).

58. A. Pasquazi, M. Peccianti, Y. Park, B. E. Little, S. T. Chu, R. Morandotti, J. Azaña, and D. J. Moss, "Sub-picosecond phase-sensitive optical pulse characterization on a chip," Nat. Photonics **5**, 618-623 (2011).

59. D. Duchesne, M. Peccianti, M. R. E. Lamont, M. Ferrera, L. Razzari, F. Légaré, R. Morandotti, S. Chu, B. E. Little, and D. J. Moss, "Supercontinuum generation in a high index doped silica glass spiral waveguide," Opt. Express **18**, 923 (2010).

60. E. Yablonovitch and T. J. Gmitter, "Photonic band structure: the face-centered-cubic case," J. Opt. Soc. Am. A **7**, 1792-1800 (1990).

61. A. Gondarenko, J. S. Levy, and M. Lipson, "High confinement micron-scale silicon nitride high Q ring resonator," Opt. Express **17**, 11366–11370 (2009).

62. L. Tong, R. R. Gattass, J. B. Ashcom, S. He, J. Lou, M. Shen, I. Maxwell, and E. Mazur, "Subwavelength-diameter silica wires for low-loss optical wave guiding," Nature **426**, 816-819 (2003).

63. M. Volatier, D. Duchesne, R. Morandotti, R. Arès, and V. Aimez, "Extremely high aspect ratio GaAs and GaAs/AlGaAs nanowaveguides fabricated using chlorine ICP etching with $N_2$-promoted passivation," Nanotechnology **21**, 134014 (2010).

64. P. Russell, "Photonic crystal fibers," Science **299**, 358-362 (2003).

65. D. Duchesne, M. Ferrera, L. Razzari, R. Morandotti, B. Little, S. T. Chu, and D. J. Moss, "Nonlinear Optics in Doped Silica Glass Integrated Waveguide Structures," in *Frontiers in Guided Wave Optics and Optoelectronics*, Bishnu Pal, ed. (2010), pp. 269-294.

66. J. Meier, W. S. Mohammed, A. Jugessur, L. Qian, M. Mojahedi, and J. S. Aitchison, "Group velocity inversion in AlGaAs nanowires," Opt. Express **15**, 12755-12762 (2007).

67. P. P. Markowicz, V. K. S. Hsiao, H. Tiryaki, A. N. Cartwright, P. N. Prasad, K. Dolgaleva, N. N. Lepeshkin, and R. W. Boyd, "Enhancement of third-harmonic generation in a polymer-dispersed liquid-crystal grating," Appl. Phys. Lett. **87**, 051102 (2005).

68. S. J. Wagner, B. M. Holmes, U. Younis, A. S. Helmy, J. S. Aitchison, and D. C. Hutchings, "Continuous wave second-harmonic generation using domain-disordered quasi-phase matching waveguides," Appl. Phys. Lett. **94**, 151107 (2009).

69. A. Fiore, V. Berger, E. Rosencher, N. Laurent, S. Theilmann, N. Vodjdani, and J. Nagle, "Huge birefringence in selectively oxidized GaAs/AlAs optical waveguides," Appl. Phys. Lett. **68**, 1320-1322 (1996).

70. L. Scaccabarozzi, M. M. Fejer, Y. Huo, S. Fan, X. Yu, and J. S. Harris, "Enhanced second-harmonic generation in AlGaAs/$Al_xO_y$ tightly confining waveguides and resonant cavities," Opt. Lett. **31**, 3626-3628 (2006).

71. S. Ducci, L. Lanco, V. Berger, A. De Rossi, V. Ortiz, and M. Calligaro, "Continuous-wave second-harmonic generation in modal phase matched semiconductor waveguides," Appl. Phys. Lett. **84**, 2974-2976 (2004).

72. D. Duchesne, K. A. Rutkowska, M. Volatier, F. Légaré, S. Delprat, M. Chaker, D. Modotto, A. Locatelli, C. De Angelis, M. Sorel, D. N. Christodoulides, G. Salamo, R. Arès, V. Aimez, and R. Morandotti, "Second harmonic generation in AlGaAs photonic wires using low power continuous wave light," Opt. Express **19**, 12408-12417 (2011).

73. A. S. Helmy, B. Bijlani, and P. Abolghasem, "Phase matching in monolithic Bragg reflection waveguides," Opt. Lett. **32**, 2399-2401 (2007).

74. L. A. Eyres, P. J. Tourreau, T. J. Pinguet, C. B. Ebert, J. S. Harris, M. M. Fejer, L. Becouarn, B. Gerard, and E. Lallier, "All-epitaxial fabrication of thick, orientation-patterned GaAs films for nonlinear optical frequency conversion," Appl. Phys. Lett. **79**, 904-906 (2001).

75. E. U. Rafailov, P. Loza-Alvarez, C. T. A. Brown, W. Sibbett, R. M. De La Rue, P. Millar, D. A. Yanson, J. S. Roberts, and P. A. Houston, "Second-harmonic generation from a first-order quasi-phase-matched GaAs/AlGaAs waveguide crystal," Opt. Lett. **26**, 1984-1986 (2001).

76. J. H. Marsh, "Quantum well intermixing," Semicond. Sci. Technol. **8**, 1136-1155 (1993).

77. K. McIlvaney, M. W. Street, A. S. Helmy, S. G. Ayling, A. C. Bryce, J. H. Marsh, and J. S. Roberts, "Selective quantum-well intermixing in GaAs-AlGaAs structures using impurity-free vacancy diffusion," IEEE J. Quantum Electron. **33**, 1784-1793 (1997).

78. S. J. Wagner, B. M. Holmes, U. Younis, I. Sigal, A. S. Helmy, S. Member, J. S. Aitchison, and D. C. Hutchings, "Difference Frequency Generation by Quasi-Phase Matching in Periodically Intermixed Semiconductor Superlattice Waveguides," IEEE J. Quantum Electron. **47**, 834-840 (2011).

79. D. C. Hutchings, "Theory of Ultrafast Nonlinear Refraction in Semiconductor Superlattices," IEEE J. Sel. Top. Quant. **10**, 1124-1132 (2004).

80. J. S. Aitchison, M. W. Street, N. D. Whitbread, D. C. Hutchings, J. H. Marsh, G. T. Kennedy, and W. Sibbett, "Modulation of the second-order nonlinear tensor components in multiple-quantum-well structures," IEEE J. Sel. Top. Quant. **4**, 695-700 (1998).

81. A. Saher Helmy, D. C. Hutchings, T. C. Kleckner, J. H. Marsh, A. C. Bryce, J. M. Arnold, C. R. Stanley, J. S. Aitchison, C. T. A. Brown, K. Moutzouris, and M. Ebrahimzadeh, "Quasi phase matching in GaAs-AlAs superlattice waveguides through bandgap tuning by use of quantum-well intermixing," Opt. Lett. **25**, 1370 (2000).

82. P. Scrutton, M. Sorel, D. C. Hutchings, J. S. Aitchison, and A. S. Helmy, "Characterizing Bandgap Gratings in GaAs : AlAs Superlattice Structures Using Interface Phonons," IEEE Photonics Technol. Lett. **19**, 677-679 (2007).

83. K. Zeaiter, D. C. Hutchings, R. M. Gwilliam, K. Moutzouris, S. Venugopal Rao, and M. Ebrahimzadeh, "Quasi-phase-matched second-harmonic generation in a GaAs/AlAs superlattice waveguide by ion-implantation-induced intermixing," Opt. Lett. **28**, 911-913 (2003).

84. David C. Hutchings, Marc Sorel, Khalil Zeaiter, Aaron Zilkie, Bert Leesti, Amr Saher Helmy, Peter Smith, and Stewart Aitchison, "Quasi-phase-matched second harmonic generation with picosecond pulses in GaAs/AlAs superlattice waveguides," in *Nonlinear Guided Waves and Their Applications* (Optical Society of America, 2004), p. TuA5.

85. K. J. Vahala, "Optical microcavities," Nature **424**, 839-46 (2003).

86. P. Del'Haye, A. Schliesser, O. Arcizet, T. Wilken, R. Holzwarth, and T. J. Kippenberg, "Optical frequency comb generation from a monolithic microresonator," Nature **450**, 1214-7 (2007).

87. T. J. Kippenberg, R. Holzwarth, and S. A. Diddams, "Microresonator-based optical frequency combs," Science **332**, 555-9 (2011).

88. Q. Lin, T. J. Johnson, R. Perahia, C. P. Michael, and O. J. Painter, "A proposal for highly tunable optical parametric oscillation in silicon micro-resonators.," Opt. Express **16**, 10596-10610 (2008).

89. T. Kippenberg, S. Spillane, and K. Vahala, "Kerr-Nonlinearity Optical Parametric Oscillation in an Ultrahigh-Q Toroid Microcavity," Phys. Rev. Lett. **93**, 18-21 (2004).

90. D. Ramirez, A. Rodriguez, H. Hashemi, J. Joannopoulos, M. Soljačić, and S. Johnson, "Degenerate four-wave mixing in triply resonant Kerr cavities," Phys. Rev. A **83**, 033834 (2011).

91. A. Rodriguez, M. Soljacic, J. D. Joannopoulos, and S. G. Johnson, "$\chi^{(2)}$ and $\chi^{(3)}$ harmonic generation at a critical power in inhomogeneous doubly resonant cavities," Opt. Express **15**, 7303-7318 (2007).

92. P. P. Absil, J. V. Hryniewicz, B. E. Little, P. S. Cho, R. A. Wilson, L. G. Joneckis, and P.-T. Ho, "Wavelength conversion in GaAs micro-ring resonators," Opt. Lett. **25**, 554-556 (2000).

93. M. Ferrera, L. Razzari, D. Duchesne, R. Morandotti, Z. Yang, M. Liscidini, J. E. Sipe, S. Chu, B. E. Little, and D. J. Moss, "Low-power continuous-wave nonlinear optics in doped silica glass integrated waveguide structures," Nat. Photonics **2**, 737-740 (2008).

94. Z. Zhang, M. Dainese, L. Wosinski, and M. Qiu, "Resonance-splitting and enhanced notch depth in SOI ring resonators with mutual mode coupling," Opt. Express **16**, 4621 (2008).

95. G. I. Stegeman, "Material figures of merit and implications to all-optical waveguide switching," Proceedings of SPIE **1852**, 75-89 (1993).

96. K. Dolgaleva, W. C. Ng, L. Qian, J. S. Aitchison, M. C. Camasta, and M. Sorel, "Broadband self-phase modulation, cross-phase modulation, and four-wave mixing in 9-mm-long AlGaAs waveguides," Opt. Lett. **35**, 4093-4095 (2010).

97. A. S. Helmy, D. C. Hutchings, T. C. Kleckner, J. H. Marsh, A. C. Bryce, J. M. Arnold, C. R. Stanley, J. S. Aitchison, C. T. A. Brown, K. Moutzouris, and M. Ebrahimzadeh, "Quasi phase matching in GaAs-AlAs superlattice waveguides through bandgap tuning by use of quantum-well intermixing," Opt. Lett. **25**, 1370-1372 (2000).

98. D. C. Hutchings, S. J. Wagner, B. M. Holmes, U. Younis, A. S. Helmy, and J. S. Aitchison, "Type-II quasi phase matching in periodically intermixed semiconductor superlattice waveguides," Opt. Lett. **35**, 1299-1301 (2010).

99. S. J. Wagner, S. C. Kumar, O. Kokabee, B. M. Holmes, U. Younis, M. Ebrahim Zadeh, D. C. Hutchings, A. S. Helmy, and J. S. Aitchison, "Performance and limitations of quasi-phase matching semiconductor waveguides with picosecond pulses," in *Proceedings of SPIE* (2010), Vol. 7750, p. 77501K.

100. S. J. Wagner, B. M. Holmes, U. Younis, I. Sigal, A. S. Helmy, J. S. Aitchison, and D. C. Hutchings, "Difference Frequency Generation by Quasi-Phase Matching in Periodically Intermixed Semiconductor Superlattice Waveguides," IEEE J. Quantum Electron. **47**, 834-840 (2011).

101. B. Bijlani, P. Abolghasem, and A. S. Helmy, "Second harmonic generation in ridge Bragg reflection waveguides," Appl. Phys. Lett. **92**, 101124 (2008).

102. U. Younis, B. M. Holmes, D. C. Hutchings, and J. S. Roberts, "Towards Monolithic Integration of Nonlinear Optical Frequency Conversion," IEEE Photonics Technol. Lett. **22**, 1358-1360 (2010).

103. Y. Okawachi, K. Saha, J. S. Levy, Y. H. Wen, M. Lipson, and A. L. Gaeta, "Octave-spanning frequency comb generation in a silicon nitride chip," Opt. Lett. **36**, 3398-3400 (2011).

104. L. Fu, V. G. Ta'eed, E. C. Mägi, I. C. M. Littler, M. D. Pelusi, M. R. E. Lamont, A. Fuerbach, H. C. Nguyen, D.-I. Yeom, and B. J. Eggleton, "Highly nonlinear chalcogenide fibres for all-optical signal processing," Optical and Quantum Electronics **39**, 1115-1131 (2008).

105. M. D. Pelusi, V. G. Ta'eed, M. R. E. Lamont, S. Madden, D.-Y. Choi, B. Luther-Davies, and B. J. Eggleton, "Ultra-High Nonlinear $As_2S_3$ Planar Waveguide for 160-Gb/s Optical Time-Division Demultiplexing by Four-Wave Mixing," IEEE Photonics Technol. Lett. **19**, 1496-1498 (2007).

106. K. K. Chow, C. Shu, C. Lin, and A. Bjarklev, "Extinction ratio improvement by pump-modulated four-wave mixing in a dispersion-flattened nonlinear photonic crystal fiber," Opt. Express **13**, 8900-8905 (2005).

107. A. Pasquazi, Y. Park, J. Azaña, F. Légaré, R. Morandotti, B. E. Little, S. T. Chu, and D. J. Moss, "Efficient wavelength conversion and net parametric gain via Four Wave Mixing in a high index doped silica waveguide," Opt. Express **18**, 7634-7641 (2010).

108. J. M. H. Elmirghani and H. T. Mouftah, "All-optical wavelength conversion: technologies and applications in DWDM networks," IEEE Communications Magazine **38**, 86-92 (2000).

109. E. Goulielmakis, V. S. Yakovlev, A. L. Cavalieri, M. Uiberacker, V. Pervak, A. Apolonski, R. Kienberger, U. Kleineberg, and F. Krausz, "Attosecond control and measurement: lightwave electronics," Science **317**, 769-775 (2007).

110. S. A. Diddams, L. Hollberg, and V. Mbele, "Molecular fingerprinting with the resolved modes of a femtosecond laser frequency comb," Nature **445**, 627-630 (2007).

111. M. J. Thorpe, K. D. Moll, R. J. Jones, B. Safdi, and J. Ye, "Broadband cavity ringdown spectroscopy for sensitive and rapid molecular detection," Science **311**, 1595-1599 (2006).

112. R. Loudon, *The Quantum Theory of Light*, 3rd ed. (Oxford University Press, USA, 2000), p. 448.

113. A. Savchenkov, A. Matsko, D. Strekalov, M. Mohageg, V. Ilchenko, and L. Maleki, "Low Threshold Optical Oscillations in a Whispering Gallery Mode $CaF_2$ Resonator," Phys. Rev. Lett. **93**, 243905 (2004).

114. L. Razzari, D. Duchesne, M. Ferrera, and R. Morandotti, "CMOS-compatible integrated optical hyper-parametric oscillator," Nat. Photonics **4**, 41-45 (2009).

115. T. Carmon, L. Yang, and K. Vahala, "Dynamical thermal behavior and thermal self-stability of microcavities.," Opt. Express **12**, 4742-4750 (2004).

116. J. Levy, A. Gondarenko, and M. Foster, "CMOS-compatible multiple-wavelength oscillator for on-chip optical interconnects," Nat. Photonics **4**, 2-5 (2009).

117. C. J. S. de Matos, J. R. Taylor, and K. P. Hansen, "Continuous-wave, totally fiber integrated optical parametric oscillator using holey fiber," Opt. Lett. **29**, 983-985 (2004).

118. M. E. Marhic, K. K.-Y. Wong, L. G. Kazovsky, and T.-E. Tsai, "Continuous-wave fiber optical parametric oscillator," Opt. Lett. **27**, 1439-1441 (2002).

119. A. Alduino, L. Liao, R. Jones, M. Morse, B. Kim, W.-Z. Lo, J. Basak, B. Koch, H.-F. Liu, H. Rong, M. Sysak, C. Krause, R. Saba, D. Lazar, L. Horwitz, R. Bar, S. Litski, A. Liu, K. Sullivan, O. Dosunmu, N. Na, T. Yin, F. Haubensack, I.-wei Hsieh, J. Heck, R. Beatty, H. Park, J. Bovington, S. Lee, H. Nguyen, H. Au, K. Nguyen, P. Merani, M. Hakami, and M. Paniccia, "Demonstration of a High Speed 4-Channel Integrated Silicon Photonics WDM Link with Hybrid Silicon Lasers," in *Integrated Photonics Research, Silicon and Nanophotonics* (Optical Society of America, 2010), p. PDIWI5.